\documentclass{aa}  

\usepackage[colorlinks=true,allcolors=blue]{hyperref}

\usepackage{threeparttable}
\usepackage[rightcaption]{sidecap}

\usepackage{graphicx}
\usepackage{txfonts}
\usepackage{xspace}
\newcommand{\kms}{km~s$^{-1}$\xspace}
\newcommand{\ergs}{erg~s$^{-1}$\xspace}
\newcommand{\Msunyr}{M$_{\odot}$~yr$^{-1}$\xspace}
\newcommand{\Msun}{M$_{\odot}$\xspace}

\newcommand{\hg}{H$\gamma$\xspace}
\newcommand{\hb}{H$\beta$\xspace}
\newcommand{\ha}{H$\alpha$\xspace}

\newcommand{\hei}{He\,{\sc{i}}\xspace}

\newcommand{\oiii}{[O\,{\sc{iii}}]\xspace}
\newcommand{\oii}{[O\,{\sc{ii}}]\xspace}

\newcommand{\sii}{[S\,{\sc{ii}}]\xspace}
\newcommand{\siii}{[S\,{\sc{iii}}]\xspace}

\newcommand{\nii}{[N\,{\sc{ii}]}\xspace}

\newcommand{\barolo}{$^{\rm 3D}${\rm Barolo}\xspace}

\newcommand{\GS}{GS18660\xspace}

\begin{document}

   \title{GA-NIFS: the highest-redshift ring galaxy candidate from a head-on collision}

   \author{Michele Perna
          \inst{\ref{iCAB}}\thanks{e-mail: mperna@cab.inta-csic.es}
          \and
          Santiago Arribas\inst{\ref{iCAB}}
          \and 
          Luca Costantin\inst{\ref{iCAB}}
          \and
          Pablo~G.~P\'erez-Gonz\'alez\inst{\ref{iCAB}}
          \and 
          Carlota Prieto-Jim\'enez\inst{\ref{iCAB},\ref{iUCM}}
          \and 
          Bruno~Rodr\'iguez~Del~Pino\inst{\ref{iCAB}}
          \and
          Francesco~D'Eugenio\inst{\ref{iKav},\ref{iCav}}
          \and
          Isabella~Lamperti\inst{\ref{iUNIFI}, \ref{iOAA}}
          \and
          Filippo~Mannucci\inst{\ref{iOAA}}
          \and
          Hannah \"{U}bler\inst{\ref{iMPE}}
          \and
          Torsten Böker\inst{\ref{iESAusa}}
          \and
          Andrew~J.~Bunker\inst{\ref{iOxf}}
          \and
          Stefano~Carniani\inst{\ref{iNorm}}
          \and
          St\'ephane~Charlot\inst{\ref{iSor}}
          \and
          Roberto~Maiolino\inst{\ref{iKav},\ref{iCav},\ref{iUCL}}
          \and
          Elena Bertola\inst{\ref{iOAA}}
          \and
          Daniel Ceverino\inst{\ref{iUAM},\ref{iCIAFF}}
          \and
          Chiara Circosta\inst{\ref{iIRAM}}
          \and
          Giovanni Cresci\inst{\ref{iOAA}}
          \and 
          Jan Scholtz\inst{\ref{iKav},\ref{iCav}}
          \and
          Giacomo~Venturi\inst{\ref{iNorm}}
          }

   \institute{
            Centro de Astrobiolog\'ia (CAB), CSIC--INTA, Cra. de Ajalvir Km.~4, 28850 -- Torrej\'on de Ardoz, Madrid, Spain\label{iCAB}
    \and
             Departamento de F\'{i}sica de la Tierra y Astrof\'{i}sica, Facultad de Ciencias F\'{i}sicas, Universidad Complutense de Madrid, E-28040, Madrid, Spain \label{iUCM}
    \and
            Kavli Institute for Cosmology, University of Cambridge, Madingley Road, Cambridge, CB3 0HA, UK\label{iKav}
    \and
            Cavendish Laboratory - Astrophysics Group, University of Cambridge, 19 JJ Thomson Avenue, Cambridge, CB3 0HE, UK\label{iCav}
    \and
            Università di Firenze, Dipartimento di Fisica e Astronomia, via G. Sansone 1, 50019 Sesto F.no, Firenze, Italy\label{iUNIFI}
    \and
            INAF - Osservatorio Astrofisico di Arcetri, Largo E. Fermi 5, I-50125 Firenze, Italy\label{iOAA}
    \and
            Max-Planck-Institut f\"ur extraterrestrische Physik (MPE), Gie{\ss}enbachstra{\ss}e 1, 85748 Garching, Germany\label{iMPE}
    \and    
            European Space Agency, c/o STScI, 3700 San Martin Drive, Baltimore, MD 21218, USA\label{iESAusa}
    \and
            Department of Physics, University of Oxford, Denys Wilkinson Building, Keble Road, Oxford OX1 3RH, UK\label{iOxf}
    \and
            Scuola Normale Superiore, Piazza dei Cavalieri 7, I-56126 Pisa, Italy\label{iNorm}
    \and
            Sorbonne Universit\'e, CNRS, UMR 7095, Institut d’Astrophysique de Paris, 98 bis bd Arago, 75014 Paris, France\label{iSor} 
    \and
            Department of Physics and Astronomy, University College London, Gower Street, London WC1E 6BT, UK\label{iUCL}  
    \and
            Departamento de Fisica Teorica, Modulo 8, Facultad de Ciencias, Universidad Autonoma de Madrid, 28049 Madrid, Spain\label{iUAM}
    \and
            CIAFF, Facultad de Ciencias, Universidad Autonoma de Madrid, 28049 Madrid, Spain\label{iCIAFF}
    \and 
            Institut de Radioastronomie Millim\'etrique (IRAM), 300 rue de la Piscine, 38400 Saint-Martin-d'H\`eres, France\label{iIRAM}
             }

   \date{Received September 15, 1996; accepted March 16, 1997}

 
  \abstract
   {Collisional ring galaxies are a rare class of interacting systems, making up only $\sim 0.01\%$ of galaxies in the local Universe. 
   Their formation is typically attributed to a head-on collision of a massive galaxy with a compact satellite (intruder), triggering density waves that, propagating outward,  produce the characteristic ring morphology. Here, we present the discovery and detailed analysis of \GS, the most distant ring galaxy known to date, at $z=3.076$, identified in JWST/NIRSpec integral field spectrograph (IFS) observations as part of the GA-NIFS programme.
   }
   {This work aims to characterise the physical and dynamical properties of \GS and shed light into the formation of its ring. Specifically, we analyse the ionized gas properties, stellar populations, and gas kinematics of the system, and use the observed geometry to constrain the timescale of the collision.}
   {Our analysis is based on JWST/NIRSpec IFS data, including low-resolution ($R\sim100$) spectroscopy covering 0.6--5.3~$\mu$m ($\sim$2000--13000~\AA\ rest-frame), and high-resolution ($R\sim2700$) spectroscopy covering 1.66--3.17~$\mu$m (4000--7700~\AA\ rest-frame).  
   Multi-wavelength techniques (e.g. emission line diagnostics, full spectral fitting, and gas dynamics)  are applied to derive nebular gas conditions and stellar population properties.
   }
   {Gas kinematic analysis reveals that \GS exhibits a rotating disk component with an additional radial expansion velocity of $\sim 200$~\kms, consistent with a propagating collisional wave. Nebular line diagnostics indicate intense star formation (SFR $\sim 100$~\Msunyr) along the ring and in the nucleus. 
   Stellar population analysis shows that the most recent star formation episode, occurring within the last $\approx 50$~Myr, predominantly took place in the ring.
   We also identify a close companion, the intruder galaxy responsible for the collision, moving away with a relative velocity of $\sim 425$~\kms. }
   {
   The evidence strongly favours a collisional origin for the ring in \GS, though the presence of a recently formed bar (and hence a resonance ring) cannot be completely excluded. This discovery provides a unique window on the physics of galaxy interactions and the rapid assembly of stellar structures in the early Universe.}

   
   \authorrunning{M. Perna et al.}

   \maketitle
%


\section{Introduction}

Collisional ring galaxies represent a rare and visually striking class of interacting galaxies, long recognised for their distinctive morphologies (e.g. \citealt{Arp1966, Dalcanton2025}). They typically form through a head-on or an off-centre collision between a compact intruder galaxy and a larger, more massive disc galaxy. This encounter triggers outwardly propagating density waves in the disc galaxy, leading to the formation of a characteristic ring-like morphology (e.g. \citealt{Lynds1976,Struck-Marcell1990}). Such systems provide unique laboratories for studying the dynamical and star-formation processes driven by galaxy interactions. 
While several examples have been identified in the nearby Universe (e.g. \citealt{Madore2009}), the advent of citizen science and machine learning applied to large-area surveys is now enabling the discovery of thousands of candidate ring galaxies at low-$z$ (e.g. \citealt{Buta2017, KrishnakumarKalmbach2024, Ulivi2025euclid, Zhang2025rings}). However, their high-redshift ($z > 0.5$) counterparts remain rare due to observational challenges, limiting our understanding of their role in early galaxy evolution.

Previous observations suggested that collisional-driven ring-like morphologies in the early Universe may have been more prevalent than in the local Universe, potentially driven by high gas fractions and frequent interactions at cosmic noon ($z\sim 1-3$; \citealt{Elmegreen2006}). However, more recent observations and simulations indicate that such systems were as rare at cosmic noon as they are today ($\sim 0.01\%$, \citealt{Elagali2018}). The most distant confirmed collisional ring galaxy to date is at $z\sim 2.2$, identified by \cite{Yuan2020NatAs}. 
More generally, ring galaxies have been observed out to $z \sim 2$, although their physical origin is not always unambiguously determined (e.g. \citealt{Genzel2014quenching,Genzel2020,NestorShachar2025}).

The advent of the James Webb Space Telescope (JWST) has significantly enhanced our ability to detect and analyse high-$z$ galaxies with unprecedented sensitivity and spatial resolution. In particular, NIRCam and NIRSpec observations enable detailed investigations of the morphology, kinematics, and star formation properties of distant interacting systems (e.g. \citealt{Nelson2023,Robertson2023,Lamperti2024, Ubler2024gn20, Parlanti2025hz4, Prieto-Jimenez2025b14}). 
Just recently, two collisional ring systems were discovered: the ``cosmic owl'' at z = 1.14 (\citealt{Li2025ring,vanDokkum2025}), and the pair UDS35606 -- UDS35616 at $z = 1.61$ (\citealt{Khoram2025}). 
In this paper, we present the discovery and detailed analysis of a collisional ring galaxy at even higher redshift, $z \sim 3$. 

The galaxy discussed in this work, known as \GS, at coordinates RA (J2000) = 03:32:54.877, DEC (J2000) = --27:45:10.40, is a star-forming galaxy at a spectroscopic redshift of $z = 3.0760$ (\citealt{Tasca2017}), with a stellar mass of log(M$_\star$/M$_\odot$) = $10.39\pm0.03$, and star formation rate log(SFR/\Msunyr) = $2.10\pm0.05$  (\citealt{Pacifici2016}). With the measured SFR and stellar mass, this object falls on the star-forming main sequence of galaxies at $z \sim 3$ (\citealt{Santini2017}). The target was specifically selected for NIRSpec integral field spectrograph (IFS) observations due to its extended morphology, as observed in HST F160W imaging, with a semi-major axis of 0.43\arcsec, corresponding to 3.4 kpc (\citealt{vanderWel2012}). This extended structure made it an ideal target for leveraging the spatially resolved capabilities of NIRSpec IFS.

While the HST imaging provided evidence for an extended structure, the ring morphology was not clearly discernible, due to the limited sensitivity and wavelength coverage (e.g. not covering \ha emission; see Appendix~\ref{sec:aHST}). In contrast, NIRSpec observations reveal a well-defined ring structure (Fig.~\ref{fig:3coloursNIRSpec}), characterized by enhanced star formation along its periphery and a dynamically complex core, consistent with theoretical models of collisional ring formation (e.g. \citealt{Renaud2018}).
Using a combination of multi-wavelength techniques applied to IFS data, we characterise its stellar population, ionised gas properties, and kinematics, providing insights into the physical conditions of this high-$z$ collisional ring galaxy. We also investigate alternative ring formation mechanisms, but we discard them on the basis of observational evidence. 

This paper is organised as follows: Section \ref{sec:observations} describes the observations. In Sect. \ref{sec:datareduction} we describe our data reduction procedures. 
In Sect. \ref{sec:environment} we present the NIRSpec colour-composite image of \GS and identify the main sources in its surroundings; 
Sect. \ref{sec:analysis} presents the data analysis; 
Sects. \ref{sec:results:kinematics} and \ref{sec:results:ism} present the kinematic and physical properties inferred from emission line analysis; 
Sect. \ref{sec:results:stellar} describes the stellar properties of \GS. 
Finally, Sect. \ref{sec:discussion} discusses the implications of our findings in the context of different ring formation mechanisms, and Sect. \ref{sec:conclusions} summarises our conclusions.
Throughout, we adopt a \cite{Chabrier2003} initial mass function (IMF, $0.1-100~M_\odot$) and a flat $\Lambda$CDM cosmology with $H_0=70$~km~s$^{-1}$~Mpc$^{-1}$, $\Omega_\Lambda=0.7$, and $\Omega_m=0.3$.
In our analysis of NIRSpec data, we use vacuum wavelengths according to their calibration.

\section{Observations}\label{sec:observations}

\GS was observed on November 26, 2024, under programme \#1216 (PI: K. Isaak), as part of the NIRSpec IFS survey `Galaxy Assembly with NIRSpec IFS' (GA-NIFS\footnote{\url{https://ga-nifs.github.io/}}, e.g. \citealt{RodriguezDelPino2024, Marconcini2024,  Scholtz2024cos30, Perna2025ct}). The project is based on the use of the NIRSpec IFS mode, which provides spatially resolved spectroscopy over a contiguous 3.1$^{\prime\prime} \times$ 3.2$^{\prime\prime}$ sky area, with a sampling of 0.1$^{\prime\prime}$/spaxel and a comparable spatial resolution (\citealt{Boker2022, Rigby2023}). 

The IFS high-spectral resolution observations were taken with the grating/filter pair G235H/F170LP. This results in a data cube with spectral resolution $R\sim2700$ over the wavelength range 1.7--3.1 $\mu$m \citep{Jakobsen2022}.
The observations were taken with the NRSIRS2 readout pattern (\citealt{Rauscher2017}) with 20 groups per integration and one integration per exposure, using a 12-point medium cycling dither pattern, resulting in a total exposure time of 4.9 hours.
Additionally, we used low ($R\sim 100$) resolution observations covering the wavelength range 0.6--5.3~$\mu$m. These were taken with the NRSIRS2RAPID readout pattern, using 26 groups per integration and yielding a total exposure time of 1.3 hours.

\section{Data reduction}\label{sec:datareduction}

We downloaded raw data files from the Barbara A. Mikulski Archive for Space Telescopes (MAST) and subsequently processed them with the JWST Science Calibration pipeline (version
1.17.1) under the recommended Calibration Reference Data System (CRDS) context jwst\_1322.pmap\footnote{\url{https://jwst-crds.stsci.edu/display_build_contexts/}}. 
All the individual raw images were first processed for detector level corrections using the Detector1Pipeline module of the pipeline (Stage1 hereinafter). Then, the individual products (count-rate images) were calibrated through Calwebb\_spec2 (Stage2 hereinafter), where wcs-correction, flat-fielding, and the flux-calibrations are applied to convert the data from units of count-rate to flux density. The individual Stage2 images were then resampled and co-added onto a final data cube through the Calwebb\_spec3 processing (Stage3 hereinafter).
To improve data quality, we applied several modifications to the default reduction steps, some of which were previously described in \citet{Perna2023}. 
Key modifications included are the following:
\begin{itemize}
    \item Zero-level correction: the individual count-rate frames were further processed at the end of Stage1, to correct for different bias levels in the dithered frames: for each image, we subtracted the median value (computed considering the entire image after masking all the 30 IFU slices and the fixed slit slices of NIRSpec at the centre of the count-rate image) to get a base level consistent with zero counts per second. 
    We note that some datasets have different bias levels in the four detector outputs\footnote{See e.g. Fig. 1 in \href{https://jwst-docs.stsci.edu/jwst-near-infrared-spectrograph/nirspec-instrumentation/nirspec-detectors/nirspec-detector-readout-modes-and-patterns/nirspec-irs2-detector-readout-mode}{JWST readthedocs}.}. 
    To account for such effect in \GS R100 data, we removed this bias for individual amplifiers per frame.
    \item 1/f noise correction: we corrected the count-rate images  through a polynomial fit at the end of stage1. 
    \item Masking of unreliable regions: pixels affected by strong cosmic rays and those associated with leakage from failed open MSA shutters were flagged as `do not use' when combining dithers. In addition, the two-pixel-wide edges of slices were masked to exclude regions with unreliable corrections in the flat-field calibration (sflat) files. Finally, we note that sflat reference files are impacted by MSA leakage, leading to inconsistencies in correction factors, where some leakage features were accounted for (but leaving significant residuals), while others were not (resulting in excessively high sflat corrections). To mitigate potential artifacts in the final products, we opted to remove these affected regions.
    \item Outlier rejection: We applied the Laplacian edge detection algorithm (\citealt{vanDokkum2001}) as implemented by \cite{DEugenio2024NatAs} to improve residual cosmic ray and artifact removal.
    \item Sub-sampling with dither and drizzle weighting: This resulted in final data cubes with 0.05\arcsec$\times$0.05\arcsec spaxels, improving spatial sampling.
\end{itemize}

We note that the noise given in the data cube (`ERR' extension) is underestimated compared to the actual noise in the data.
Following the method introduced in our previous works (e.g. \citealt{Ubler2023}), we re-scaled the ‘error’ vector in each spaxel to match the noise estimated from the standard deviation of the continuum in spectral regions free of line
emission. We applied a correction factor of $\sim 1.6$. 
Extending our previous works, this is achieved by normalizing the error vector using its running median and then scaling it by the running standard deviation of the continuum in line-free regions, computed as 1.48 times the median absolute deviation (e.g. \citealt{Lamperti2020}), which is less affected by outliers and provides more reasonable uncertainties. See also \citet{DEugenio2025LRDz5} for a similar approach.


\subsection{R100 wavelength calibration}

Previous studies have shown that R100 spectra are typically redshifted by a few hundred \kms with respect to medium- and high-resolution spectra, both in NIRSpec IFS and MSA observations (e.g. \citealt{Perez-Gonzalez2024, Bunker2024dr,DEugenio2025dr3}). In  Appendix~\ref{sec:R100offset}  we demonstrate that R100 data in NIRSpec IFS mode are affected by a systematic wavelength offset of $24_{-3}^{+5}$~\AA~ (observer-frame) relative to R2700.  A wavelength calibration correction should be applied to R100 data to ensure accurate velocity measurements in cases where an absolute wavelength reference is not available.

\subsection{Additional data processing steps}

No background exposures were included in these observations, so
we performed a background subtraction by extracting a mean spectrum from a large manually-selected, signal-free area in each of the R2700 and R100 cubes. This mean background spectrum was then modelled with a polynomial function, and finally we subtracted the best-fit model from each spaxel.

As part of our general data reduction strategy, we tested additional procedures that are not required for the \GS\ dataset but may be of broader interest for the community. 
A detailed description of these procedures is provided in Appendix~\ref{app:extraDR}.

\section{Source environment}\label{sec:environment}

\GS, despite being located in the GOODS-S field and previously observed as part of CANDELS, is not covered by any of the JWST NIRCam programmes focused on this field (e.g. \citealt{Oesch2023, Williams2023}). In particular, \GS lies just outside the NIRCam imaging footprints of the JADES ultra-deep imaging programme (\citealt{Eisenstein2023}). 

\begin{figure}[!t]
   \centering
   \includegraphics[width=0.48\textwidth]{{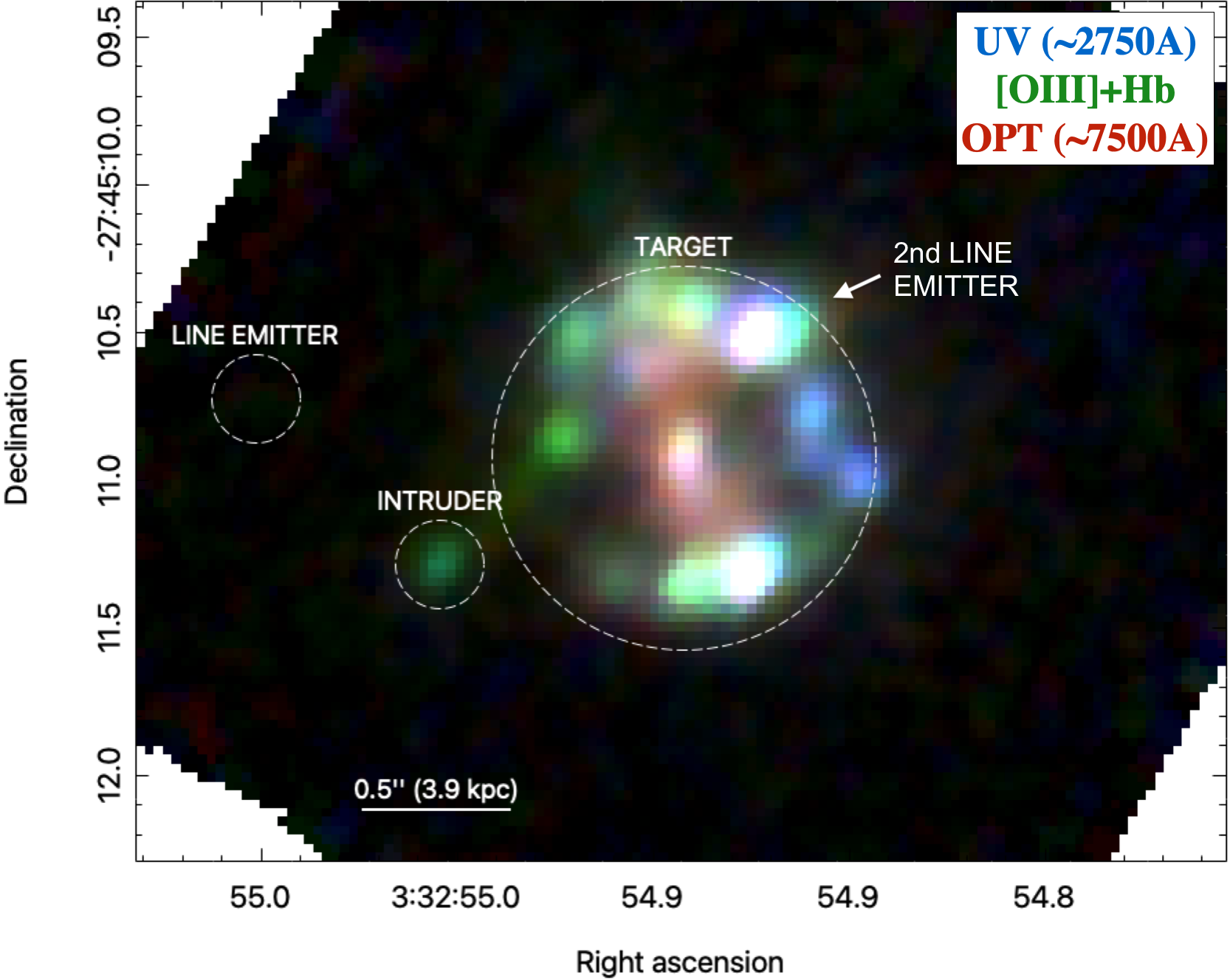}}
   \caption{Colour-composite image of GS18660 (the `target') and its close companions (the `intruder' and the `line emitters'), derived from NIRSpec R100. Blue shows the rest-frame UV, green the \oiii-\hb, and red the optical continuum, as labelled. The circles mark the regions used to extract the integrated spectra analysed in Sect. \ref{sec:environment}. The arrow marks a very faint source detected in \oiii and \ha (see also Fig.~\ref{fig:R2700MapsTarget}, bottom).}
    \label{fig:3coloursNIRSpec}%
    \end{figure}

Figure \ref{fig:3coloursNIRSpec} presents the NIRSpec IFS three-colour image of \GS, where rest-frame UV continuum emission ($\sim 2750~\AA$) is shown in blue, \oiii\!$\lambda\lambda$4960,5008 and \hb emission in green, and rest-frame optical continuum ($\sim 7500~\AA$) in red. This visualisation allows us to identify two main systems within the FOV. \GS appears as an extended source with a bright nuclear region and a prominent ring composed of several clumps that show distinct colours. The nucleus and a bar-like structure orientated along the north-south direction are dominated by optical continuum emission. 
Additionally, a compact satellite, labelled `intruder' and located $\sim 0.85\arcsec$ south-east of the \GS nucleus, is visible as a green clump, indicating strong \oiii emission. Doubly ionised oxygen emission also connects the main galaxy to the intruder.


%

   \begin{figure*}[!h]
   \centering
   \includegraphics[width=0.95\textwidth, trim=0mm 10mm 0mm 2mm,clip]{{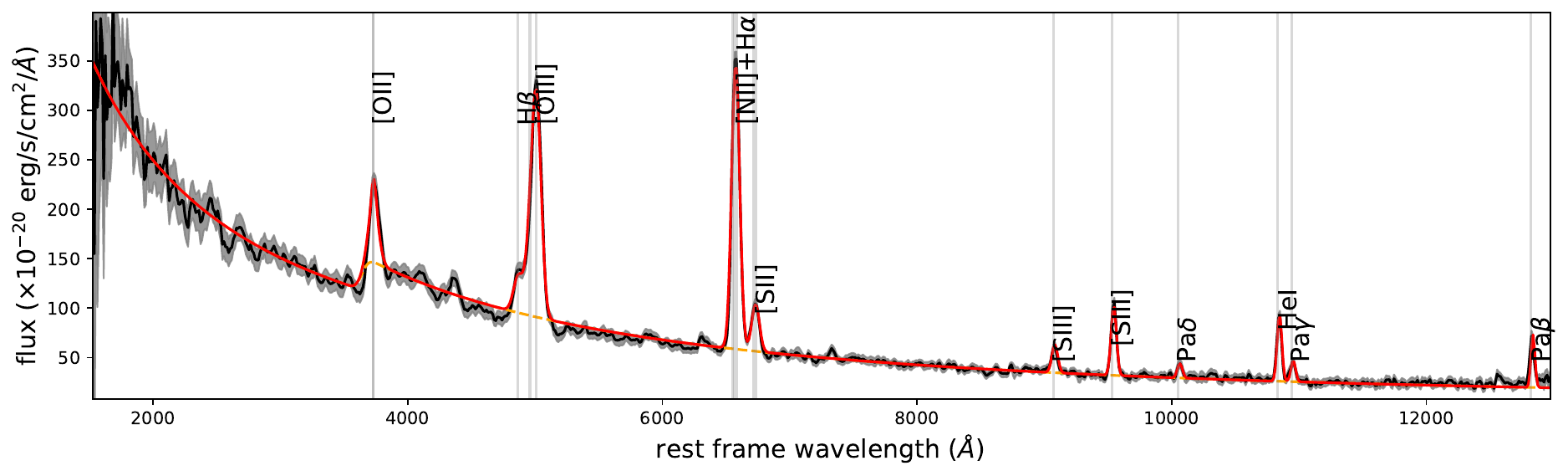}}
    \includegraphics[width=0.945\textwidth, trim=0mm 02mm 0mm 2mm,clip]{{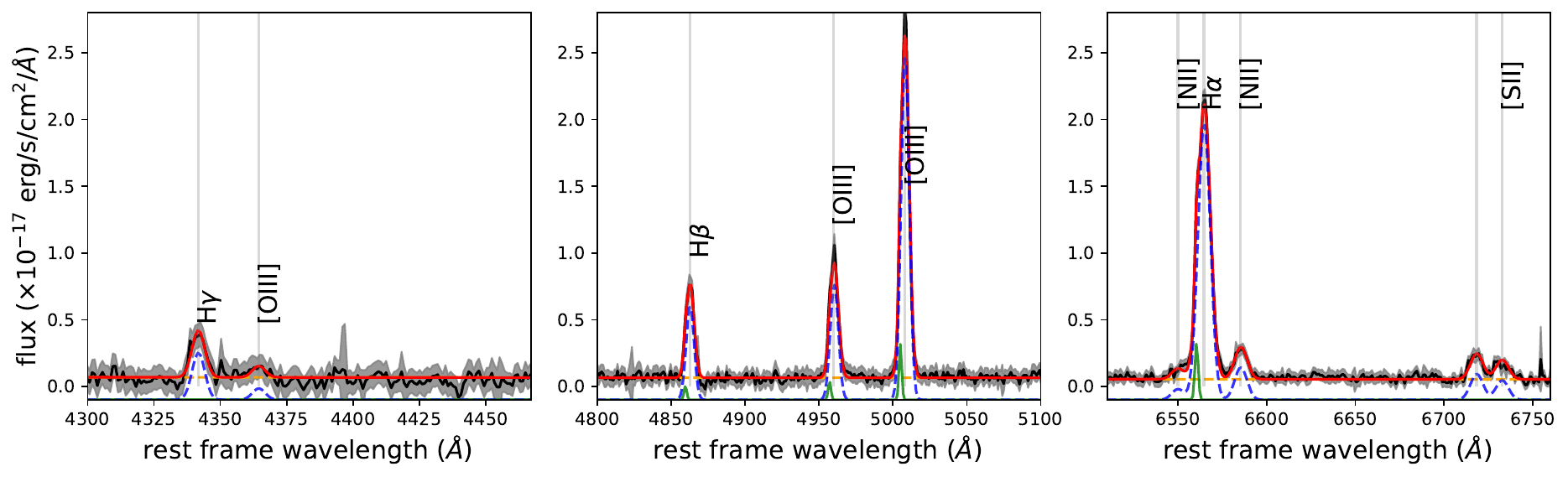}}

    \caption{Integrated spectra of \GS, with best-fit results. Top panel: The black curve represents the R100 spectrum  obtained integrating over the entire extension of \GS ($r=0.65\arcsec$), with 3$\sigma$ uncertainties in light-grey. The red curve denotes the best-fit model, including both emission lines and continuum, while the dashed orange line represents the continuum alone. Fitted emission lines are marked with grey vertical lines and labelled. Bottom panels: R2700 spectrum integrated over the same region (black curves, with 3$\sigma$ uncertainties in grey) and best-fit results in the vicinity of modelled emission lines: \hg and \oiii\!$\lambda$4363 (left); \hb and \oiii\!$\lambda\lambda$4960, 5008 (centre); \ha, \nii and \sii doublets (right). 
    }
    \label{fig:integratedspectraGS18660}%
    \end{figure*}
%

   \begin{figure*}[!h]
   \centering
   \includegraphics[width=0.95\textwidth, trim=0mm 10mm 0mm 5mm]{{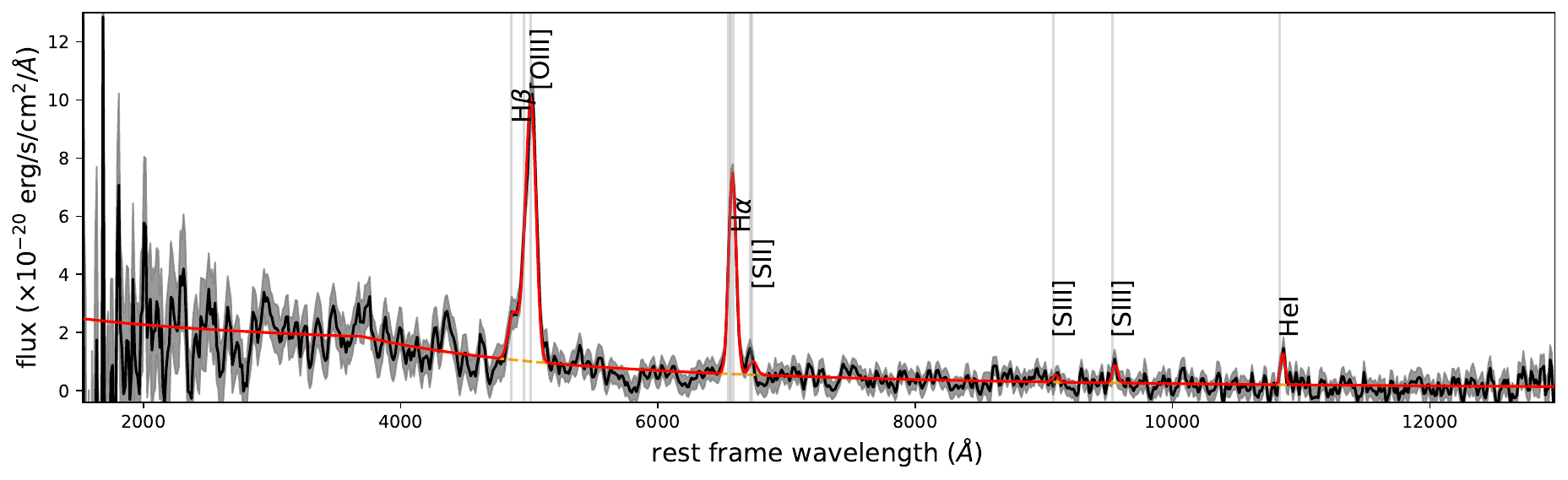}}
    \includegraphics[width=0.91\textwidth, trim=0mm 02mm 0mm 2mm,clip]{{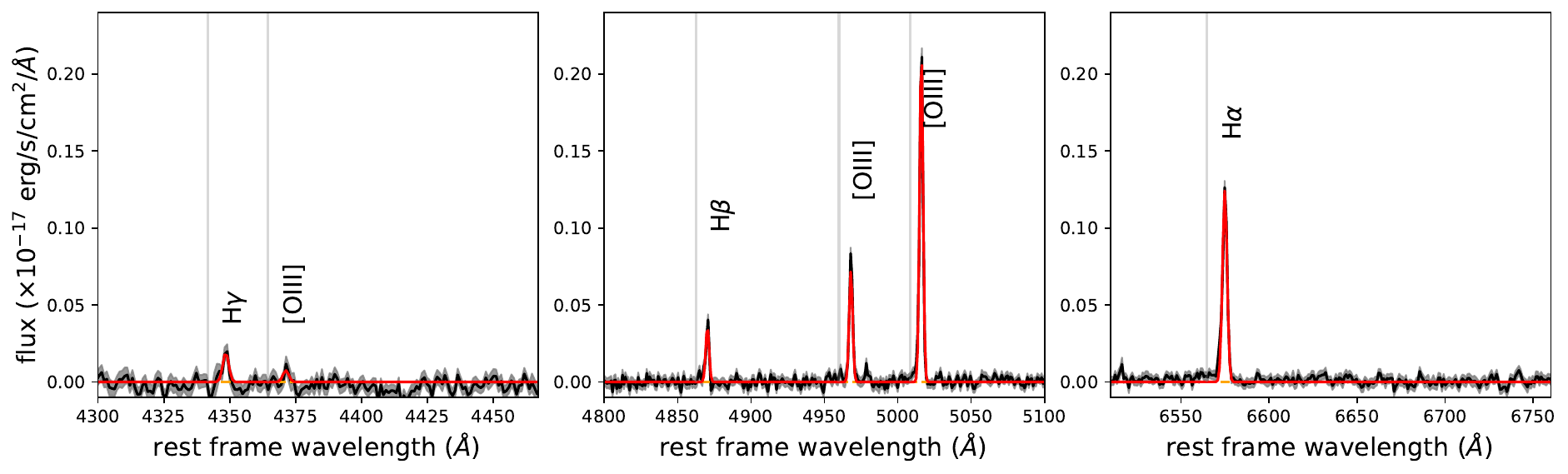}}

    \caption{Integrated R100 and R2700 spectra of the intruder galaxy, within a $r=0.15\arcsec$ aperture. See Fig.~\ref{fig:integratedspectraGS18660} for further details. 
    The \sii doublet is likely undetected, as the observed peak in R100 is offset from its expected position 
    and may instead result from continuum fluctuations. 
    }
    \label{fig:integratedspectraINTRUDER}%
    \end{figure*}

Figure \ref{fig:integratedspectraGS18660} shows the integrated spectra of \GS, extracted within a radius of 0.65\arcsec\ (large dashed circle in Fig.~\ref{fig:3coloursNIRSpec}). The top panel displays the R100 spectrum over the full wavelength range covered by NIRSpec, while the bottom panels zoom in on key optical emission lines also covered by the R2700 data. Both spectra are shown with their respective best-fit continuum and line models, obtained using the methodology described in Sect.~\ref{sec:analysis}. From these models, we determine the systemic redshift of \GS to be $z = 3.0766\pm 0.0001$.  

Figure \ref{fig:integratedspectraINTRUDER} presents the R100 and R2700 spectra of the intruder, extracted from a circular aperture with a radius of 0.15\arcsec. This source is significantly fainter than \GS, with only a few emission lines detected, including \hg, \hb, \ha, and \oiii lines in the R2700 spectrum. The R100 data also reveal \siii\!$\lambda$9533 and \hei 1.08~$\mu$m, in addition to \ha, \hb, and the \oiii doublet seen in R2700. The continuum emission is detected only in the R100 spectrum. From the high-resolution R2700 lines, we measure a systemic redshift of $z = 3.08239$, indicating that the intruder is redshifted by $\sim 425$~\kms relative to \GS.




A detailed investigation of the NIRSpec cubes reveals another faint line emitter at $\sim 1.5\arcsec$ ($\sim 12$~kpc) east of \GS (see circle in Fig.~\ref{fig:3coloursNIRSpec}). This emitter has a compact morphology and is detected in \oiii and \ha (both in R100 and R2700), but not in the continuum nor in the \oiii+\hb pseudo-narrow-band image shown in Fig.~\ref{fig:3coloursNIRSpec}. Figure \ref{fig:integratedspectraLINEEMITTER} shows the high-resolution spectrum of this target, for which we derive a redshift $z = 3.07678$, hence with a velocity offset of just $\sim 10$~\kms from \GS. Additionally, we identified a further compact line emitter on the north-west side of the ring, with a velocity offset of $\sim-$370~\kms relative to \GS (see Fig.~\ref{fig:3coloursNIRSpec}, and Sect.~\ref{sec:kinandfluxdistribution}). 

In summary, NIRSpec IFS data allowed us to discover a compact satellite (the intruder) and two line emitters in the vicinity of the main galaxy. 
These findings suggest that \GS may reside in a small galaxy group, a scenario commonly observed in collisional galaxies at lower redshift (e.g. \citealt{Wong2017}) and predicted by cosmological simulations (\citealt{Elagali2018}). Alternatively, the proximity and kinematics of the line emitters could indicate that they are remnants or tidal features associated with a recent or ongoing merger involving the intruder. 

\section{Data analysis}\label{sec:analysis}

\subsection{Emission line fit of R2700 data}
\label{sec:analysisR2700}

In this section, we describe the method used to derive the emission line properties by fitting the R2700 data. The method applies to both individual-spaxel and spatially-integrated spectra. 

We first modelled the continuum level by fitting the flux across the entire wavelength range, excluding
the spectral regions within $\pm 1200$~\kms from the position of the emission lines. We modelled the continuum flux using low-order polynomials. 
After subtracting the continuum, we fitted the most prominent gas emission lines by using the Levenberg-Marquardt least-squares fitting code CAP-MPFIT (\citealt{Cappellari2017}). 
In particular, we modelled the \ha and \hb lines, the \oiii\!$\lambda\lambda$4960,5008, \nii\!$\lambda\lambda$6550,85, and \sii\!$\lambda\lambda$6718,32 doublets with a combination of Gaussian profiles, applying a simultaneous fitting procedure, so that all line features of a given kinematic component have the same velocity centroid and FWHM (e.g. \citealt{Perna2020}).  Moreover, the relative flux of the \nii and \oiii doublet components was fixed to 2.99; the \sii flux ratio was required to be within the range $0.44 <
f(\lambda 6718)/ f(\lambda 6732) < 1.42$ (\citealt{Osterbrock2006}). Each fit was performed with one and two Gaussian components. The final number of kinematic components used to model individual spectra was derived on the basis of the Bayesian information criterion (\citealt{Schwarz1978}). 
Figures~\ref{fig:integratedspectraGS18660} and~\ref{fig:integratedspectraINTRUDER} (bottom panels) illustrate examples of the fitting results, with  two kinematic components for the integrated spectrum of the target and a single component for the intruder.

\subsection{Emission line fit of R100 data}
\label{sec:analysisR100}

In this section, we describe the method used to derive the emission line properties by fitting the R100 data. 
Due to their coarse line spread function (LSF; \citealt{Boker2023}), lines are frequently blended. To accurately fit these data, it is necessary to construct a high-resolution model and convolve it with the LSF (e.g. \citealt{Jones2024HFLS3}). While the R100 data are not well suited for detailed kinematic analysis or precise redshift determinations, their broad wavelength coverage enables the simultaneous observation of many emission lines such as \oii$\!\lambda\lambda3727,30$, \siii\!$\lambda\lambda 9071,9533$, \hei 1.08~$\mu$m not covered by the G235H grating,  alongside the continuum emission from $\sim 0.2$ to $\sim 1.2~\mu$m (rest-frame).

To fit the spectra, we first model the continuum emission following the approach of \cite{Jones2024HFLS3}, which we summarise here. Spectral regions within $\pm 5000$~\kms of the emission lines were excluded from the fit. The spectrum of \GS exhibits a pronounced Balmer break around 3645 Å and distinct power-law behaviours in the UV and optical regions (see Figs.~\ref{fig:integratedspectraGS18660} and \ref{fig:integratedspectraINTRUDER}). We therefore parametrise the continuum as follows.
The UV region ($\lambda < 3645 \AA$) is modelled with a power-law, $f_\lambda \propto \lambda^{\beta_{\rm UV}}$, where $\beta_{\rm UV}$ is the slope of the UV continuum.
The optical region ($\lambda > 3645 \AA$) is modelled with a separate power-law function, $f_\lambda \propto  \lambda^{\beta_{\rm opt}}$, with $\beta_{\rm opt}$ representing the optical continuum slope.
A discontinuity at $3645 \AA$ is introduced to account for the Balmer break between the UV and optical power-law components. To properly account for the instrumental broadening, the discontinuity is convolved with the NIRSpec LSF. 

Following the continuum fitting, we model the main emission lines 
with a single Gaussian component. In addition to the constraints introduced in Sect.~\ref{sec:analysisR2700} for \oiii and \nii, we further constrain the \oii doublet line ratio within the range 0.35--1.50 taking into account the electron density dependence (e.g. \citealt{Pradhan2006}); for the \siii lines  we adopted a theoretical ratio
of \siii\!$\lambda$9532/$\lambda$9069 = 2.5 (\citealt{Vilchez1996}).  Given that the spectral resolution at R100 is insufficient to resolve velocity substructures, we adopt single Gaussian profiles for each line. 

We note that the spectral resolution of NIRSpec IFS R100 data appears to be higher than pre-flight expectations (see e.g. \citealt{Jones2024HFLS3, Lamperti2024}). \citet{Marshall2025pearls} analysed R100 observations of the planetary nebula SMP LMC 58 and found that the observed emission line profiles were narrower than predicted by the nominal LSF. Their study determined a revised spectral resolving power of R~=~$1.14_{-0.09}^{+0.11} \times~$R$_{\rm pre-flight}$.
To ensure consistency in our analysis, we applied this rescaling, which resulted in a good agreement between the intrinsic velocity dispersion of emission lines in R100 and R2700 for the integrated spectrum of \GS.

\begin{table}[h]
\centering
\caption{Spatially integrated properties from the fit analysis.}%
\begin{tabular}{|lc|}
\hline

\hline
\GS (target): & \\
\hline

z & $3.07662_{-0.00009}^{+0.00005}$ \\ 
$\sigma_v$ & $150_{-5}^{+15}$~\kms\\
L(\oiii) & $(1.59\pm 0.01)\times 10^{43}$~\ergs \\
L(\ha) & $(1.61\pm 0.01)\times 10^{43}$~\ergs \\
\ha/\hb & $3.84_{-0.07}^{+0.11}$\\
E(B--V) & $0.31\pm 0.01$\\
log(\oiii/\hb) & $0.582_{-0.003}^{+0.008}$ \\
log(\nii/\ha) & $-0.96_{-0.01}^{+0.12}$\\
log(\oiii5008/6343) & $1.51_{-0.05}^{+0.07}$\\
\sii 6716/6731 & $1.37\pm 0.05$\\
$\beta_{\rm UV}$, $\beta_{\rm opt}$ & $-1.24_{-0.03}^{+0.05}$, $-1.62_{-0.02}^{+0.03}$\\

\hline\hline
Intruder: & \\
\hline
z & $3.08239\pm0.00006$\\
$\sigma_v$ & $44\pm 1$~\kms\\
L(\oiii) & $(5.45\pm0.02)\times 10^{41}$~\ergs\\
L(\ha) & $(3.51\pm 0.03)\times 10^{41}$~\ergs\\
\ha/\hb & $3.92_{-0.08}^{+0.07}$\\
E(B--V) & $0.31\pm 0.04$\\
log(\oiii/\hb) & $0.78\pm 0.01$ \\
log(\nii/\ha) & $-1.45_{-0.06}^{+0.10}$\\
\oiii5008/6343 & $1.47_{-0.03}^{+0.14}$\\
$\beta_{\rm UV}$, $\beta_{\rm opt}$ & $-0.32_{-0.05}^{+0.06}$, $-2.05_{-0.03}^{+0.04}$\\

\hline\hline
Line emitter: & \\
\hline
z & $3.07678\pm 0.00003$\\
$\sigma_v$ & $20\pm 1$~\kms \\
L(\oiii) & $(5.5\pm 0.2)\times 10^{40}$~\ergs\\
L(\ha) & $(3.3\pm 0.2)\times 10^{40}$~\ergs\\
\ha/\hb & $>+4.40$\\
E(B--V) & -- \\
log(\oiii/\hb) & $>+0.86$\\
log(\nii/\ha) & $<-0.67$\\
\oiii5008/6343 & $>+1.07$\\

\hline
\hline

\end{tabular} 
\tablefoot{\GS properties  integrated over $r~=~0.65\arcsec$. Intruder and line emitter integrated over $r~=~0.15\arcsec$. Colour excess measured from \ha/\hb 
(Sect.~\ref{sec:extinction}). Luminosities not corrected for extinction.}   
\label{tab:integratedproperties}
\end{table}



\subsection{Stellar population synthesis fit}

Stellar population synthesis modelling was performed on the $R=100$ cube with the \textsc{synthesizer-AGN} code \citep{2003MNRAS.338..508P,2008ApJ...675..234P,2024ApJ...968....4P}. We 
assumed two stellar populations, each one described by a Star Formation History (SFH) following a delayed exponential function. We run models with a variety of timescales, ranging from 1~Myr (i.e. an instantaneous burst) to 1~Gyr (i.e. constant SFH). The stellar emission for each burst is described by the \citet{Bruzual2003} models, assuming a \citet{Calzetti2000} IMF with stellar masses between 0.1 and 100~$\mathrm{M}_\odot$. The stellar population ages are allowed to vary from 1 Myr up to the age of the Universe at the redshift of the galaxy. Based on the emission line ratios measured with R2700 data, we considered stellar emission models with metallicities in the range $Z_\star \in[0.2,0.4]\,Z_\odot$. Nebular line and continuum emission is modelled via Cloudy models \citep{Ferland+17}, assuming $Z_g=0.25$~$Z_\odot$, a variety of electron density ranging from $10^2$ to $10^4$\,cm$^{-3}$, and an ionization parameter range $\log(U)\in[-3.0,+0.0]$. Dust attenuation is modelled with the \citet{Calzetti2000} law, with A$_{\rm V}$ values ranging from 0 to 3~mag for each population, considered to have completely independent attenuations. The stellar population modelling was run on a spaxel-by-spaxel basis after applying a relative photometric calibration correction following the wavelength-dependent function published in \citet{DEugenio2024NatAs}.
We assumed floor relative photometric errors of 5\% and 10\%, added in quadrature to the uncertainties derived from the noise of the background, which was measured in the cube in an empty region. We also run models after binning each spectrum to reach $S/N\geq5$. The three different executions produced similar results, and in the remaining of this paper we show results for the 5\% floor relative photometric error. 

In addition, we performed independent stellar population analysis using \textsc{pPXF} (Appendix~\ref{app:ppxf}) and \texttt{CIGALE} (Appendix~\ref{app:cigale}), which yields consistent results with those from \textsc{synthesizer-AGN}. Minor differences are observed in the detailed shape of the stellar mass assembly history from \textsc{pPXF}, as discussed in Sect.~\ref{sec:results:stellar}.

   \begin{figure*}
   \centering
   \includegraphics[width=0.91\textwidth, trim=0mm 4mm 0mm 4mm]{{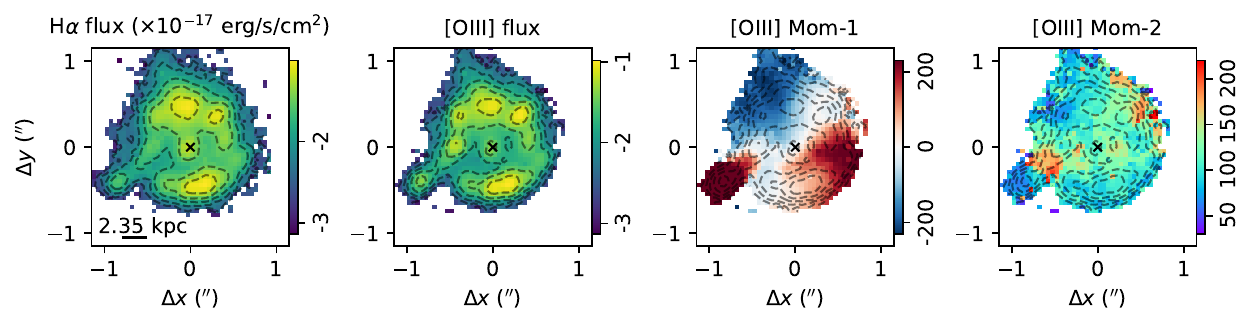}}
   \caption{ \ha and \oiii flux distributions, along with the \oiii moment-1 (velocity) and moment-2 (velocity dispersion) maps. The flux maps reveal the morphology of \GS, highlighting its ring structure and nuclear region, as well as the elongated morphology of the intruder. The velocity map displays a rotation-like pattern, with an additional high-velocity redshifted component associated with the intruder. The velocity dispersion map shows a relatively uniform distribution, with localized enhancements near the intruder, and in the western part of the ring. The contourns in the first panel trace \ha emission; those in the remaining panels trace \oiii emission. All images are oriented north up, with east to the left.}
              \label{fig:R2700MapsTotal}%
    \end{figure*}
%

   \begin{figure*}
   \centering
   \includegraphics[width=0.91\textwidth, trim=0mm 8mm 0mm 2mm,clip]{{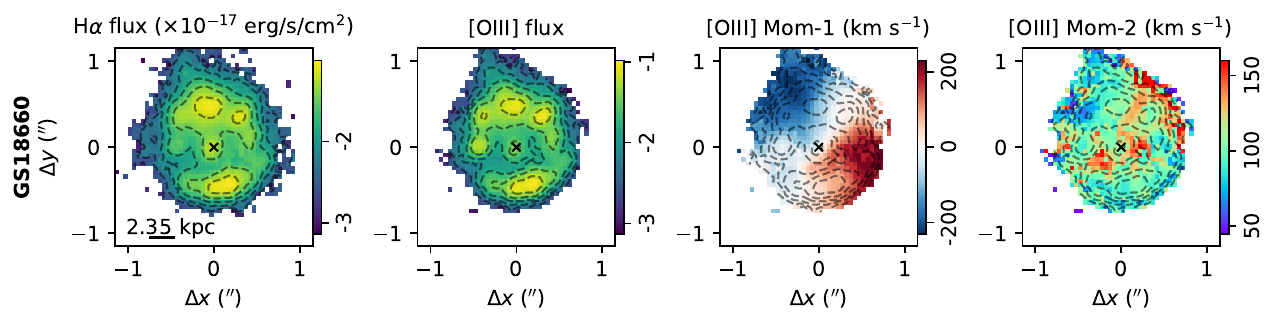}}
   \includegraphics[width=0.91\textwidth, trim=0mm 2mm 0mm 8mm,clip]{{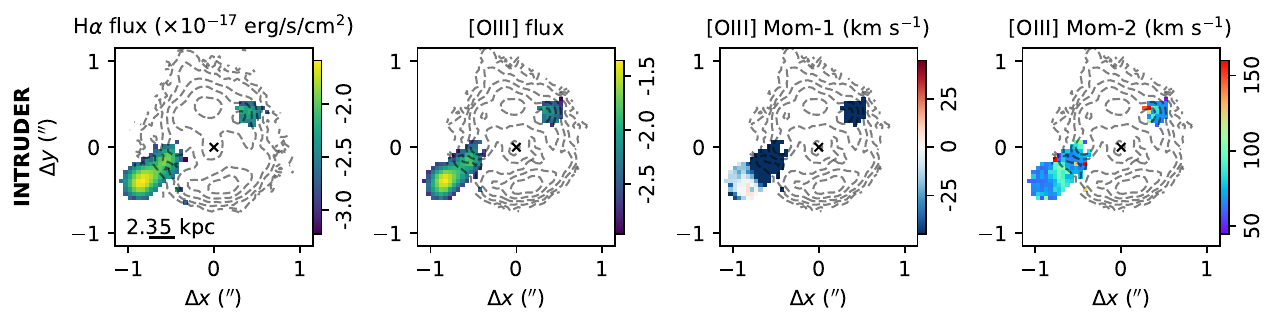}}

   \caption{\GS (top) and intruder (bottom) \ha and \oiii flux distributions, along with \oiii moment-1 and moment-2 maps. These maps are similar to those in Fig.~\ref{fig:R2700MapsTotal}, but were generated after removing all Gaussian components not associated with the \GS system (top) or the intruder (bottom). The intruder  Moment-1 map is computed with respect to its systemic velocity ($\delta v = -430~$\kms relative to \GS).}
              \label{fig:R2700MapsTarget}%
    \end{figure*}
%

%

\section{Results: gas kinematics}\label{sec:results:kinematics}

\subsection{Kinematics and emission line flux distributions}\label{sec:kinandfluxdistribution}

Figure~\ref{fig:R2700MapsTotal} presents the flux distributions of \oiii and \ha, together with the velocity and velocity dispersion maps derived from the spaxel-by-spaxel analysis. The flux maps reveal prominent clumps within the ring structure, and the intruder with an elongated morphology in the southeast region of the FOV aligned towards the central region of \GS. 
We define the nucleus of \GS as the position corresponding to the central peaks of the \ha and \oiii emission, which also coincides with the kinematic centre of the rotating disk discussed in Sect.~\ref{sec:barolo}.
Given the strong similarity between the \oiii and \ha velocity maps, we show only the \oiii velocity and velocity dispersion maps.

The velocity map exhibits a gradient from --220~\kms to +220~\kms along the direction north-east to south-west, indicating a rotational pattern. The intruder is characterized by extreme redshifted velocities, reaching up to $\sim 450$~\kms. The velocity dispersion map shows relatively high values across the entire \GS system, peaking at $\sim 200$~\kms in the vicinity of the nucleus and in the northwestern part of the ring. Notably, these peaks do not coincide with the nucleus itself. Similarly high velocity dispersions are observed near the intruder due to the presence of multi-peak line profiles, where both the target and the intruder contribute to emission at significantly different velocities along the same line of sight.

To separate the intruder from \GS, we used the multi-Gaussian decomposition of the emission-line profiles presented in Sect. \ref{sec:analysisR2700}. Components with velocity shifts in the range $-250 < \Delta v/({\rm km~ s^{-1}}) <250$ are assigned to \GS, while those outside of this range are assigned to the intruder. Given the strong velocity gradient of the intruder, we further refine this separation by assigning all positive-velocity components to the intruder if they are located in the bottom-left quadrant of the map ($\Delta x < -0.25\arcsec$). This approach yields a clearer distinction between the two systems. The resulting \oiii, \ha, and velocity maps for both \GS and the intruder are shown in Fig.~\ref{fig:R2700MapsTarget}. 

The flux and kinematic maps of \GS now exhibit a more coherent ring structure and a smoother rotational velocity pattern, confirming that contamination from the intruder was affecting the initial maps. Meanwhile, the intruder appears as an extended source with a bright core located $\sim 7$~kpc from the nucleus of \GS and an elongated bridge-like structure stretching toward the \GS nucleus. 
The intruder velocity map, shifted by +425~\kms relative to \GS systemic velocity in the figure, reveals internal velocity variations from --180 to +30~\kms (i.e. 245 to 455~\kms with respect to \GS) along the $\sim 5$~kpc-scale structure, with the smallest  velocity offsets with respect to the systemic redshift of \GS closest to the ring of the main galaxy. Additionally, we identified another compact line emitter on the north-west side of the ring, with a velocity offset of $\sim -370$~\kms relative to \GS. 

The velocity dispersion remains enhanced in the intruder spaxels, suggesting a composite nature: a compact galaxy with an elongated tidal tail formed through interaction with \GS might explain our fit results. We also note that the compact core of the intruder shows a velocity gradient along the north-east to south-west direction, with a peak-to-peak velocity of only $\sim 50$~\kms, possibly due to rotation.

   \begin{figure}
   \centering
   \includegraphics[width=0.49\textwidth, trim=0mm 0mm 0mm 6mm,clip]{{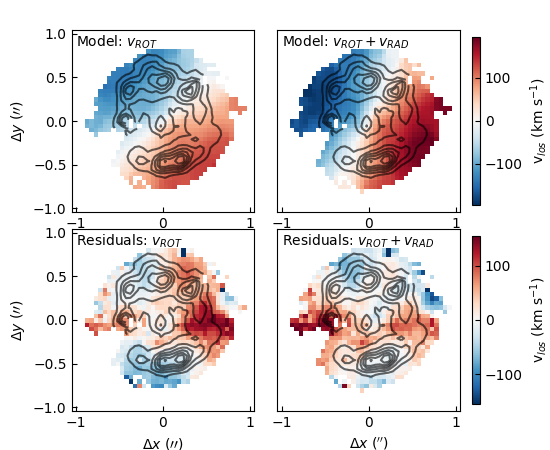}}
   \caption{Best-fit \barolo models of the \oiii velocity field in \GS. The top panels show the modelled moment-1 maps including only rotational motions (left) and both rotational and radial motions (right). The bottom panels display the corresponding residual velocity fields (data~$-$~model). The inclusion of a radial velocity component leads to a reduction in the residuals, although significant residuals remain along the east-west direction, suggesting the presence of complex gas dynamics not fully captured by the simple disc model.}
              \label{fig:3dbarolo}%
    \end{figure}

\subsection{\barolo disc modelling}\label{sec:barolo}

To test whether the gas kinematics of \GS is compatible with a rotationally supported system, we modelled the \oiii best-fit data cube with \barolo (\citealt{Diteodoro2015}). 
Specifically, we modelled the data obtained after subtracting the additional kinematic components associated with the intruder and north-west line emitter.
The main assumption of the \barolo model is that all the emitting material of the galaxy is confined to a geometrically thin disk and its kinematics are dominated by pure rotational motion. The possible presence of residual components associated with radial motions (e.g. due to outflows, inflows) can be modelled assuming a radial velocity component in addition to the rotation. 

\barolo best-fit results have been obtained following a three-step strategy, similar to the ones described in \citet{DiTeodoro2021} and \citet{Perna2022}. First, we tried different azimuthal models spanning a range of disk inclination angles $i$ with respect to the observer (5$^\circ$ to 85$^\circ$ spaced by 5$^\circ$, with 0$^\circ$ for face-on); during this step, the $i$ parameter is fixed, and the fitting minimization is performed considering the following free parameters: the rotation velocity, $v_{rot}$, the velocity dispersion, $\sigma$, and the major axis PA, $\phi$. The disk centre is fixed to the position of the nucleus, corresponding to the \ha peak. We therefore inferred the disk inclination angle considering the best-fit configuration with the minimal residuals, defined using the Eqs. 2 and 3b in \citet{Diteodoro2015}. 
Then, we run \barolo with a local flux normalisation, letting it minimise the $v_{\rm rot}$, $\sigma$, $\phi$ and $i$ parameters. In this second step, the inclination angle $i$ is left free to vary by a few degrees around the best-fit value defined in the previous step ($i = 20^\circ$). 
The best-fit parameters obtained are as follows: $i \approx 20^\circ$, $\phi\approx ~214^\circ$, $v_{\rm rot}\approx 320~$\kms, and $\sigma \approx 80$~\kms at the position of the ring, with all parameters showing radial dependences as detailed: the velocity dispersion decreases with radius from $\sim 100$ to $\sim 50$~\kms outside the ring, $v_{\rm rot}$ increases up to $\approx 300$~\kms before the ring where it goes up to $\sim 500$~\kms, the inclination is stable at $\sim 20^\circ$ and the position angle changes from 200$^\circ$ (inner regions) to 230$^\circ$ (external regions). The inclination
agrees with the value of $20^\circ$ derived by
modelling the galaxy isophotes (e.g. \citealt{Jedrzejewski1987, MendezAbreu2018, Costantin2018}).

The best-fit model obtained in this way provides a good overall reproduction of the velocity field of \GS, but significant residuals with velocities up to $150$~\kms are found along the east-west direction, hence broadly close to the minor axis ($\phi \sim 300^\circ$), and towards the north and north-west (Fig. \ref{fig:3dbarolo}, left). 
These residuals could be associated to radial flows (e.g. \citealt{vanderKruit1978, Price2021}); therefore, we performed a third \barolo run with a radial component in addition to the rotational motions. In this third run, all the other parameters were initialised to the previously determined values, and we minimise only the two velocity parameters, fixing the major axis PA to $\phi = 214^\circ$. The best fit results obtained from the third run are reported in Fig. \ref{fig:3dbarolo} (right). The inclusion of a radial component improves the fit ($\Delta \chi^2_{red}=1.5$, with final $\chi^2_{red}=1.6$), and provides smaller residuals towards north-west but it does not remove the ones along the east-west direction. This suggests that while radial motions can be present in the system, the motions in \GS are much more complex and cannot be fully reproduced with this simple model. The presence of tidal interactions and the bridge towards the companion, not perfectly removed with our multi-Gaussian fit decomposition, may also contribute to the non-perfect modelling. 
Finally, we note that warps could also contribute to the observed residuals; however, they are not required by the second-step fit and, in any case, cannot account for the pronounced asymmetries seen in the kinematics of \GS (see Fig.~1 in \citealt{DiTeodoro2021}).

The final best-fit parameters obtained from the third run are as follows: $i \approx 20^\circ$, $\phi\approx ~220^\circ$, $v_{\rm rot}\approx 320~$\kms, $\sigma \approx 80$~\kms, and $v_{\rm rad}\approx -200~$\kms at the position of the ring; small radial dependences are observed in all parameters, but the only significant ones are observed for the rotational velocity, which increases from $\sim 120$~\kms in the nuclear regions to $\gtrsim 350$~\kms outside the ring.


The value of radial velocities derived by \barolo alone does not inform whether the motion is directed inwards (inflow, compaction) or outwards (outflow, expansion). The direction of the rotation of the disc is needed to correctly interpret the nature of the radial motion. By definition of \barolo, a positive (negative) $v_{\rm rad}$ is tracking expanding (contracting) gas if the galaxy rotates clockwise and contracting (expanding)
when it rotates counter-clockwise. 
Current data do not allow us to understand if \GS rotates clockwise or counter-clockwise; we note however that collisional ring galaxies may present expansion motions at the position of the ring, and compaction in the inner regions (e.g. Fig. 19 in \citealt{Elagali2018} and Fig. 9 in \citealt{Chen2018ApJ864}). The quality of our data (in terms of spectral and angular resolution, as well as S/N) does not allow for the identification of such variations as a function of the radius. If we interpret the negative radial velocity as due to the expansion of the ring, the \GS disc would rotate counter-clockwise.

\section{Results: ISM physical properties}\label{sec:results:ism}

The combination of R100 and R2700 data allows for a detailed investigation of ISM physical properties. The R100 spectrum covers continuum emission from $\sim 0.2$ to $\sim 1.2~\mu$m rest-frame and all main optical emission lines including the \oii doublet and \siii doublet which are not covered by R2700. The high-resolution data allows us to resolve key emission lines such as \hb, \ha, and \oiii lines avoiding blending issues in R100. The \sii doublet is analysed in both resolutions: R2700 for line de-blending and R100 for total flux measurements (see Sect.~\ref{sec:ionisation}). These emission lines provide key diagnostics for gas-phase metallicity, electron density and temperature, ionisation conditions, and ionization parameter.

Since some of these diagnostics rely on faint emission lines that are often undetected in individual spaxels, we also analyse integrated spectra of distinct star-forming clumps. To identify these clumps, we apply a dendrogram analysis using the astrodendro package, which decomposes the emission-line distribution into a hierarchical structure of branches and leaves, with leaves representing the smallest unresolved line-emitting elements (\citealt{Rosolowsky2008}). Dendrograms have been widely used to study the hierarchical structure of star-forming regions (e.g. \citealt{Nayak2016, Feltre2025, Zhou2025dendro}).

Figures \ref{fig:spectraleaves} and \ref{fig:spectraleavesR100} display the R2700 and R100 spectra, respectively, with best-fit models extracted from the identified leaves (overlaid on the \oiii emission map in the right panels). The best-fit physical properties derived for each region are summarised in Table~\ref{tab:integratedproperties}.
In the following sections, we explore the properties of the ISM at the spaxel and individual leaf levels.

\begin{figure*}[ht!]
\centering
\begin{minipage}{0.692\textwidth}
  \includegraphics[width=\linewidth, trim=0mm 0mm 0mm 0mm,clip]{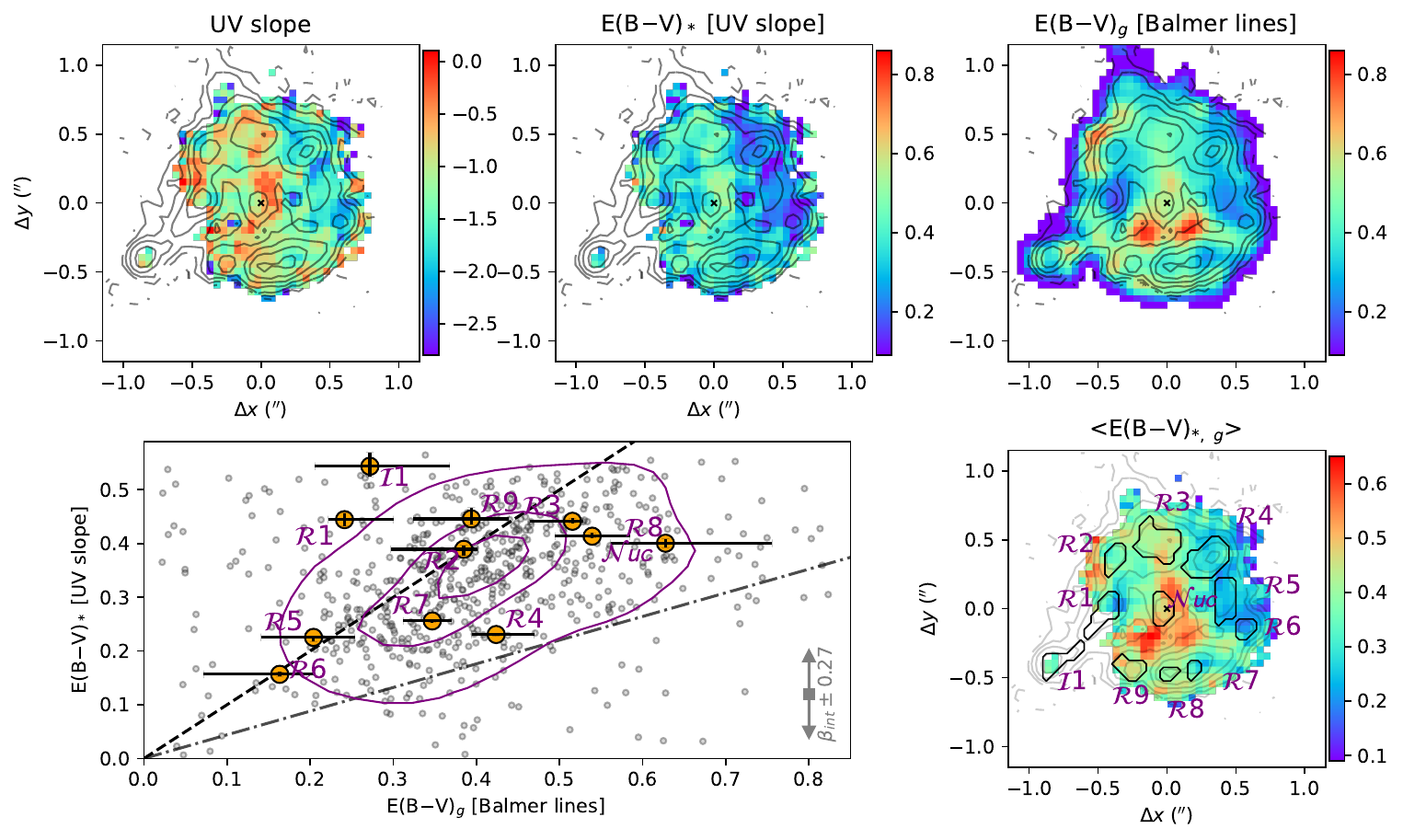}
\end{minipage}%
\hfill
\begin{minipage}{0.308\textwidth}
  \caption{Maps of UV slope and dust extinction, and gas versus stellar extinction diagram. Maps from top-left to bottom right: UV slope, stellar colour excess derived from UV slope, gas colour excess derived from the \ha/\hb, and average E(B--V) obtained combining the second and third maps. The bottom-left diagram shows a spaxel-by-spaxel comparison between the two E(B--V); the contours include 20, 68, and 90\% of the measurements, and show that most of the values are consistent, regardless a significant scatter from the 1:1 relation (dashed line; dot-dashed line indicates the \citealt{Calzetti2000} 0.44:1 relation). This scatter is likely due to measurement uncertainties as well as intrinsic deviations possibly due to differences in the attenuation of stars and gas. Large circles refer to the values from the integrated regions (leaves) shown in the bottom-right map. In the bottom-right corner of the plot, we show the effect of $\pm10\%$ variations in the assumed intrinsic UV slope. 
}\label{fig:dustextinction}
\end{minipage}
\end{figure*}

\subsection{Dust extinction}\label{sec:extinction}

To quantify dust extinction in our study, we employ two complementary methods: the Balmer decrement (\ha/\hb) and the UV spectral slope ($\beta_{\rm UV}$). The former provides insights into the attenuation affecting ionised gas regions, while the UV slope offers information on the dust extinction impacting the stellar continuum. Utilising both approaches allows for a comprehensive assessment of dust attenuation effects within the observed galaxies.

For the Balmer decrement approach, we assume a \citet{Cardelli1989} attenuation law and a fixed temperature of 15000~K. Although there could be in principle a temperature range in the FOV, the expected variation in colour excess of ionised gas E$_{\rm g}$(B--V) is $\pm0.07$ for a reasonable temperature range between 5000~K to $20000$~K, typical of the conditions in HII regions (\citealt{Osterbrock2006}). As shown in Sect.~\ref{sec:electrontemperature}, the electron temperatures in individual integrated regions can be directly measured using \oiii~$\lambda$4363, and range from 15000~K to more than 20000~K, albeit with significant uncertainties (see Table~\ref{tab:integratedproperties}). Therefore, adopting a fixed temperature of 15000~K is justified, corresponding to an intrinsic \ha/\hb ratio of 2.79, as computed with PyNeb (\citealt{Luridiana2015}). 
In order to avoid aperture losses from PSF sampling that could affect the Balmer decrement estimates, the spatially-resolved measurements are obtained from spatially integrated regions of $3\times3$ spaxels.

We also analyse the UV slope $\beta$ as an indicator of extinction.
A redder UV slope generally indicates more dust extinction, and consequently a higher E(B-V) (e.g. \citealt{Meurer1999}). 
Following \cite{Chisholm2022}, we estimate the stellar colour excess E$_*$(B--V) using the relation:

\begin{equation}
E_\star(B-V) = \frac{\log_{10} \left(\frac{\lambda_1}{\lambda_2} \right)}{ 0.4 (k_{\lambda_1} - k_{\lambda_2})} \left( \beta_\mathrm{int}^{<\lambda_1,~\lambda_2>} - \beta_\mathrm{obs}^{<\lambda_1,~\lambda_2>}\right),
\end{equation}

where $\lambda_1$ and $\lambda_2$ correspond to  1990~$\AA$ and 3500~$\AA$, respectively, aligning with the wavelength range of our NIRSpec R100 data used to measure $\beta$. $k_{\lambda_1}$ and $k_{\lambda_2}$ are the attenuation values at the two wavelengths of interest, obtained from the assumed \cite{Calzetti2000} extinction law commonly adopted for stellar continuum.  $\beta_{obs}^{<\lambda_1,~\lambda_2>}$ is the UV slope measured within the same wavelength range; $\beta_{int}^{<\lambda_1,~\lambda_2>}$ is the intrinsic one, therefore not affected by obscuration. The latter quantity can be derived from stellar population analysis, or assumed following previous works. For simplicity, we adopted the same intrinsic value measured in \cite{Chisholm2022} from the median value of 89 star-forming galaxies at $z\sim 0.3$ and consistent with the expected intrinsic stellar continuum slope obtained from Starburst99 models (\citealt{Leitherer1999}): $\beta_{int}^{<\lambda_1,~\lambda_2>} = -2.71$. Despite the slight difference in wavelength ranges  (1300--1800~$\AA$ in their work, compared with our 1990-3500~$\AA$), the UV slope is expected to remain relatively constant (see e.g. Fig. 9 in \citealt{Leitherer1999}). 

Figure \ref{fig:dustextinction} shows the UV slope and colour excess maps derived from our analysis (top panels). E$_\star$(B--V) is compared with the one obtained from the Balmer line ratio (E$_g$(B--V)) in the bottom-left panel; this plot shows that stellar and nebular gas measurements are broadly consistent, with values from 0.1 to 0.8 mag (with the Balmer decrement values which tend to overestimate the colour excess by $\lesssim 0.1~$dex). Taking into account that these values are obtained with a series of assumptions, and that $\beta$ slope and Balmer decrement can be uncertain, we assume that the two tracers are equivalent, and measure the mean E(B--V) averaging the stellar and gas estimates; the ${\rm <E(B-V)>}$ is reported in the bottom right panel of Fig.~\ref{fig:dustextinction}. A similarity between stellar and gas extinction is also supported by the spatially-integrated colour excess measurements E$_\star$(B--V) and E$_g$(B--V) we derived from the identified leaves, also reported in Fig. \ref{fig:dustextinction} (bottom-left) with orange symbols.
We note that different studies assume different intrinsic UV slopes; for example, \citet{Boquien2022} adopt $\beta_\mathrm{int}=-2.62$ for $z\sim5$ galaxies based on alternative stellar population models; in the bottom-left panel of Fig.~\ref{fig:dustextinction} we therefore show the effect of $\pm10\%$ variations in $\beta_\mathrm{int}$.

Intrinsic differences in the attenuation of stars and gas are expected: \cite{Calzetti2000} found an average value of ${\rm E_\star(B-V)/E_g(B-V)} = 0.44$ in the local Universe. Although there is no consensus in literature, there is  growing evidence that at $z \gtrsim 1$ the discrepancy between the nebular and the stellar attenuation could be smaller (e.g. ${\rm E_\star(B-V)/E_g(B-V)} \sim 0.9$ in \citealt{Puglisi2016}). Our results point in the same direction. 
In particular, the spatially integrated analysis suggests that regions dominated by young stellar populations along the ring are closer to the $1:1$ relation, while regions hosting a mixture of old and young stars (e.g. the nucleus) are more consistent with the canonical $1:0.44$ relation (see also \citealt{Paspaliaris2025}), in line with the stellar population analysis presented in Sect.~\ref{sec:results:stellar}. However, the large uncertainties on the derived quantities preclude firm conclusions.

The average colour excess map shows higher dust extinction in the circum-nuclear regions, and in the north-east external parts of the ring. A dust lane in the bar-shaped structure along the north-south direction might also be present, although current results tend to point towards a patchy distribution around the nucleus. 
In particular, the dust-lane-like structure appears more prominent on the western side of the nucleus, which may be the far side, in line with the counter-rotating disc scenario discussed in Sect.~\ref{sec:barolo}. 
However, higher angular resolution data might be required to confirm the presence of dust lanes. Overall, the ring is bluer than the rest of the galaxy, consistent with the B--V colour profiles measured by \cite{Romano2008} for a sample of nine nearby collisional ring galaxies.

\subsection{Ionisation conditions}\label{sec:ionisation}

\begin{figure}
   \centering
   \includegraphics[width=0.49\textwidth, trim=0mm 2mm 0mm 2mm,clip]{{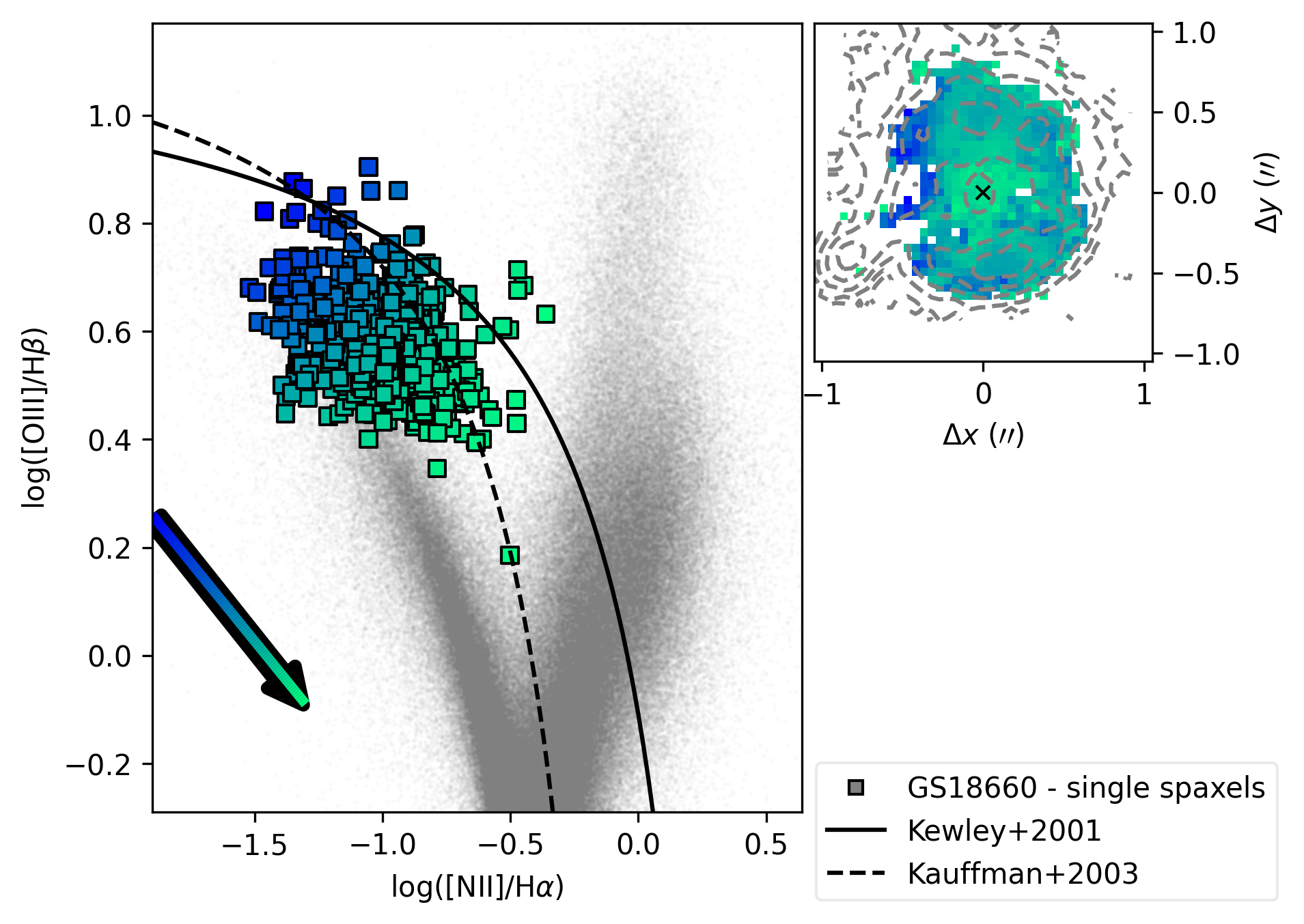}}
   \caption{Standard BPT diagnostic diagram. The colour-coded squares show single-spaxel measurements associated with the spatial regions shown in the top-right panel; green-to-blue colours mark line ratio variations, as indicated with the arrow in the bottom-left part of the diagram; only spaxels with S/N $> 3$ in all lines are shown. The solid (\citealt{Kewley2001}) and dashed (\citealt{Kauffmann2003}) curves commonly used to separate purely star-forming galaxies (below the curves) from AGN (above the curves) are also reported. Finally, grey points indicate local galaxies from SDSS (\citealt{Abazajian2009}).}
              \label{fig:BPT}%
    \end{figure}

We investigated the dominant ionisation source for the emitting gas across the \GS host using
the classical ``Baldwin, Phillips \& Terlevich'' (BPT) diagram (\citealt{Baldwin1981}). 
Figure~\ref{fig:BPT} shows the spatially resolved flux ratio diagnostics, colour-coded from green to blue to indicate increasing (decreasing) values along the x- (y-)axis of the BPT diagram. For clarity, we omit the spatially integrated measurements of individual clumps (dendrogram leaves), as they occupy a relatively narrow region of the diagram similar to the spaxel-by-spaxel values.   
For comparison, the BPT also displays other optical line ratio measurements from low-$z$ galaxies (small grey points), and the demarcation lines used to separate galaxies and AGN at $z\sim 0$. 

Spatially resolved line ratios are similar to the ones observed in other high-$z$ star-forming galaxies (e.g. \citealt{Newman2014, Perna2018lbt,Lamperti2024,Jones2025}). Notably, the nuclear regions exhibit lower \oiii/\hb\ and higher \nii/\ha\ ratios compared to the outer regions and the ring. This trend likely reflects radial variations in both metallicity and ionisation conditions, as suggested by photoionisation models (e.g. \citealt{Kewley2013}). A similar spatial pattern has also been observed in the Cartwheel galaxy (see Fig. 6 in \citealt{Ditrani2024}), and interpreted as the result of older stellar populations in the nucleus and younger, star-forming populations in the ring.

To further investigate the ionisation structure, we derived the ionisation parameter $U$, defined as the ionising photons flux per hydrogen atom density, divided by the speed of light. We adopted the \siii 9069,9532/\sii 6717,31 ratio, which is less sensitive to metallicity than the commonly used \oiii/\oii ratio (\citealt{KewleyDopita2002}). Both \siii and \sii fluxes were measured in R100, to avoid flux calibration differences between R100 and R2700 cubes (e.g. \citealt{Arribas2024}).
Line fluxes were corrected for extinction using the average $E(B-V)$ values derived in Sect.~\ref{sec:extinction}, and $U$ was computed using the empirical relation from \citet{Diaz2000}. The resulting map of ionisation parameter is shown in the right panel of Fig.~\ref{fig:ZandU}, revealing enhanced ionisation in the ring and lower values toward the nucleus, consistent with the above interpretation based on the BPT. No bar-like structure is identified on the map. For context, the left panel of Fig.~\ref{fig:ZandU} shows the corresponding metallicity distribution (see Sect.~\ref{sec:metallicity} for details), which peaks in the nuclear region.

\subsection{Electron density}\label{sec:electrondensity}

The measurement of the electron density ($n_e$) is challenging due to the faintness of the [\ion{S}{ii}]$\lambda\lambda$6716,31 lines, which prevents spatially resolved $n_e$ maps from being produced. Nevertheless, we extracted integrated spectra for the high-S/N clumps defined by the dendrogram leaves in Sect.~\ref{sec:results:ism}, and we used the [\ion{S}{ii}] doublet ratio in combination with the PyNeb package to estimate the electron density in these regions.
The derived values span a range between $n_e = 100$ and $1000$ cm$^{-3}$ (Table~\ref{tab:integratedproperties}), slightly higher than those of other ring galaxies at lower redshift. For example, \citet{Fogarty2011} reported electron densities between 150 and 300 cm$^{-3}$ in the star-forming ring of Arp 147, while the Cartwheel Galaxy displays densities in the 10–200 cm$^{-3}$ range in the ring (\citealt{Zaragoza-Cardiel2022}).

Although limited statistics preclude a robust correlation analysis, this result could be consistent with previous findings linking higher $n_e$ to regions of elevated star formation rate (e.g. \citealt{Arribas2014, Kaasinen2017}). In fact, the SFR in the \GS ring is a few times higher than that for instance measured in the local Cartwheel Galaxy (SFR$_{\rm tot}\sim 9$~\Msunyr, \citealt{Ditrani2024}, while individual clumps in \GS can have up to 30~\Msunyr).
In summary, the electron densities in \GS appear elevated compared to lower-redshift analogues, possibly reflecting more compact and intense star-forming regions in this early-Universe system.



\subsection{Electron temperature}\label{sec:electrontemperature}

We measured the electron temperature ($T_e$) in \GS\ using the [\ion{O}{iii}]$\lambda$4363 auroral line, which is detected in a subset of our leaf-integrated spectra. As with the $n_e$ measurements, we employed PyNeb to convert the [\ion{O}{iii}]$\lambda$4363/($\lambda$4959+$\lambda$5007) ratio into electron temperatures.

In the northern part of the system, particularly in regions $\mathcal{I}$1 and $\mathcal{R}$1--$\mathcal{R}$4 (see e.g. Fig.~\ref{fig:dustextinction}), we obtain $T_e$ values in the range of 1.3--1.9 $\times 10^4$~K. For the remaining regions, where the auroral line is undetected, we estimate lower limits of $T_e \sim 1.5$–$2.4 \times 10^4$~K (Table~\ref{tab:integratedproperties}). These values are elevated but not unusual for galaxies at $z > 2$, especially in regions of intense star formation or low metallicity.
For comparison, \citet{Fosbury1977} derived $T_e$ values in the range 1.4--1.8 $\times 10^4$~K in the Cartwheel Galaxy (see also \citealt{Zaragoza-Cardiel2022}), consistent with our measurements in \GS. 

Our results demonstrate that auroral-line detections and electron temperature diagnostics are feasible with NIRSpec IFS even at $z = 3$: while \oiii 4363 has now been routinely detected even in spatially-integrated spectra of $z \gtrsim 7$ galaxies with JWST (e.g. \citealt{Laseter2024, Ubler2024zs7, Sanders2025}), our study proves that $T_e$ can also be derived in individual bright clumps of star formation (see also \citealt{Marconcini2025cr7}).



\subsection{Gas metallicity}\label{sec:metallicity}

The spatially resolved study of the ISM metallicity is a key diagnostic for identifying regions of a galaxy that have undergone significant chemical enrichment through star formation, as well as those potentially influenced by the inflow of metal-poor gas from the circumgalactic medium and satellites. Taking advantage of the high S/N detection of multiple emission lines across the field, we constrain the gas-phase oxygen abundance in \GS and its close companion using well-established empirical calibrations (e.g. \citealt{Curti2017}).
 
\begin{figure}
   \centering
   \includegraphics[width=0.49\textwidth, trim=2mm 0mm 2mm 0mm,clip]{{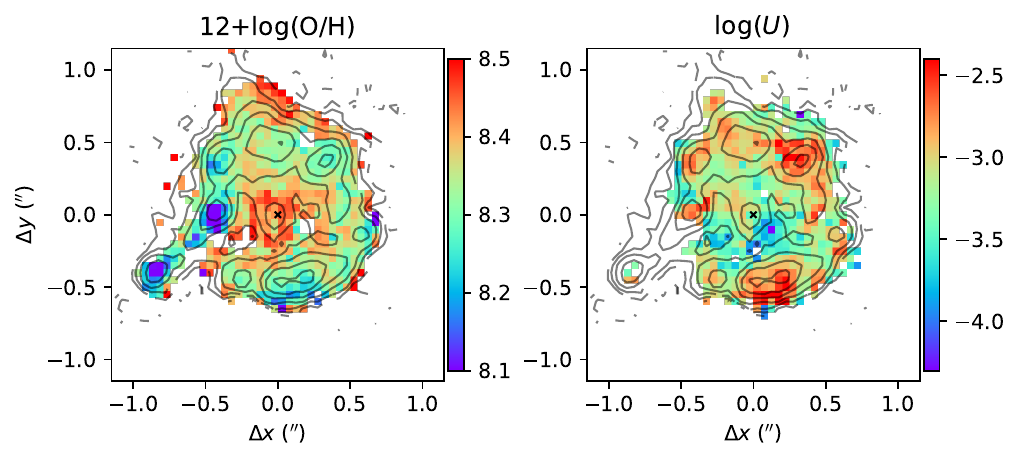}}
   \caption{Gas metallicity and ionisation parameter maps. The former is obtained from the parameters R2, R3, R23, N2, and O32 (Sect.~\ref{sec:metallicity}); the latter is derived from \siii/\sii line ratios (Sect.~\ref{sec:ionisation}). }
              \label{fig:ZandU}%
    \end{figure}

After correcting for dust attenuation (average E(B--V) map in Fig.~\ref{fig:dustextinction}), we use flux ratios between specific emission lines that correspond to standard metallicity indicators: R2 (\oii\!$\lambda\lambda$3727, 30/\hb), R3 (\oiii\!$\lambda$5008/\hb), O32 (\oiii\!$\lambda$5008/\oii\!$\lambda\lambda$3727, 30), R23 (\oii\!$\lambda\lambda$3727, 30 + \oiii\!$\lambda\lambda$4960, 5008)/\hb), and N2
(\nii\!$\lambda$6585/\ha). 
The metallicity calibrations of \citet{Curti2017} for the parameters R2, R3, R23, N2, and O32 were fitted together to generate the oxygen abundance map in Fig.~\ref{fig:ZandU} (see \citealt{Curti2024} for details).

The resulting map reveals enhanced oxygen abundances (12 + log(O/H)$\approx 8.4$–$8.5$) in the nuclear region of \GS and in the northern extension, possibly corresponding to a tidal feature (see also Sect.~\ref{sec:results:stellar}). These values are consistent with the expectations from the mass-metallicity relation (e.g. \citealt{Sanders2021}). Slightly lower values ($\sim$8.3) are found across most of the ring. Notably, the eastern star-forming clump in the ring, the bridge connecting to the intruder, and the intruder itself show even lower metallicities (12 + log(O/H)$\approx 8.1$–$8.2$). The low metallicity of the intruder is consistent with its lower stellar mass ($\approx 10^8$~\Msun, see Sect.~\ref{sec:results:stellar}), while the metal dilution in the bridge and nearby ring regions may be a tail of low-metallicity gas left behind by the intruder. 

In Table~\ref{tab:integratedproperties}, we report metallicity measurements derived from the integrated spectra of the high-S/N clumps identified as leaves in Sect.~\ref{sec:results:ism}. These integrated values are generally consistent with the spatially resolved results shown in Fig.~\ref{fig:ZandU}, reinforcing our conclusions. We note that for some integrated spectra important emission lines such as \hb, \nii, \oii are detected at low significance; this explains the upper limit in the metallicity of the intruder, and some missing measurements.   

The relatively low metallicity in the ring, despite ongoing intense star formation, can be naturally explained in the framework of collisional ring formation: when the density wave reaches the outer disc, it may trigger star formation in gas that has not been previously significantly enriched by earlier stellar generations 
(e.g. \citealt{Mapelli2012, Kouroumpatzakis2020, Zaragoza-Cardiel2022}).




\section{Results: stellar properties}\label{sec:results:stellar}

\subsection{Stellar mass}\label{sec:stellarmass}

\begin{figure}[!t]
   \centering  \includegraphics[width=\linewidth, trim=0mm 3mm 0mm 0mm,clip]{{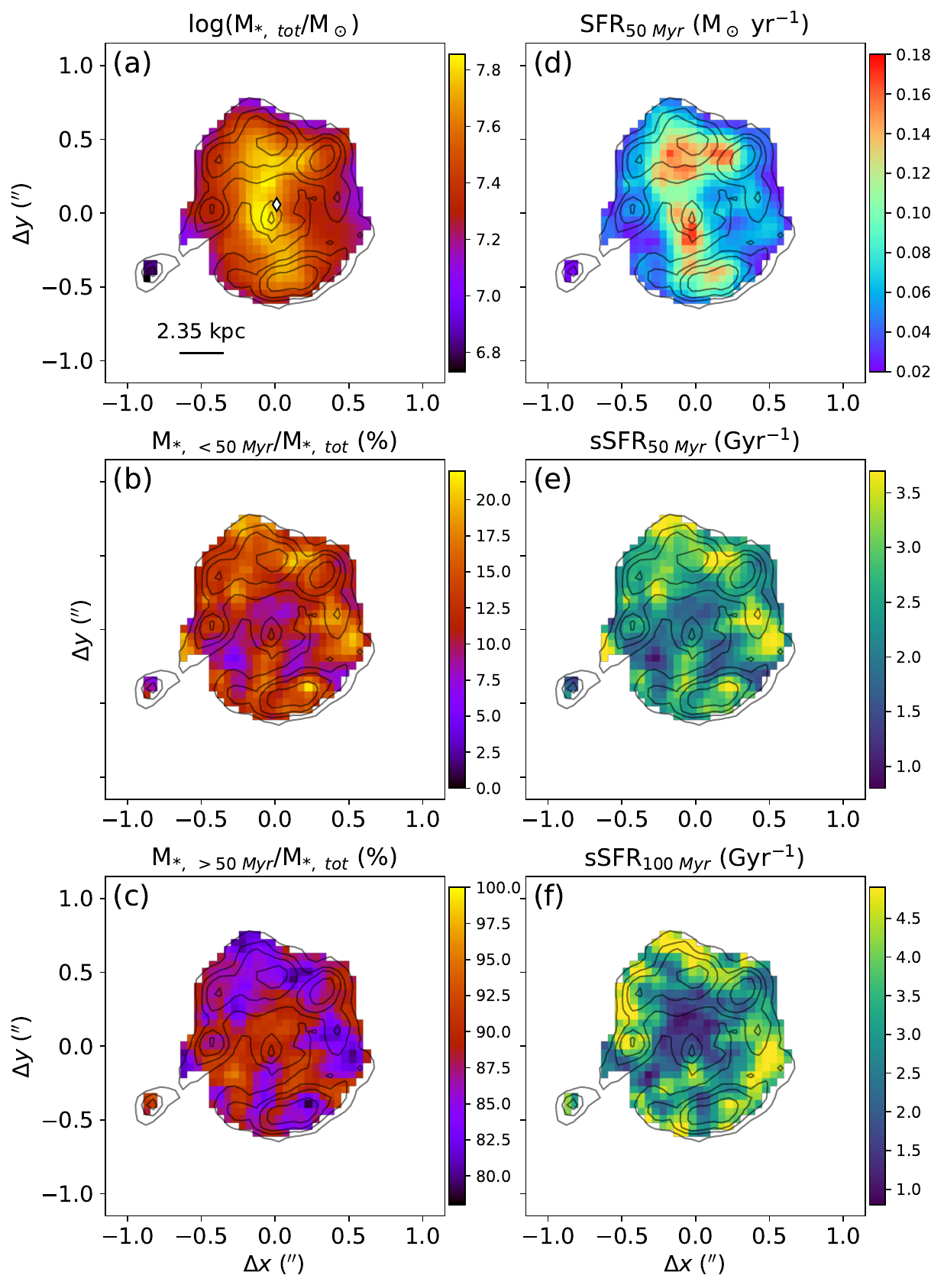}}
   \centering  \includegraphics[width=\linewidth, trim=0mm 0mm 0mm 2mm,clip]{{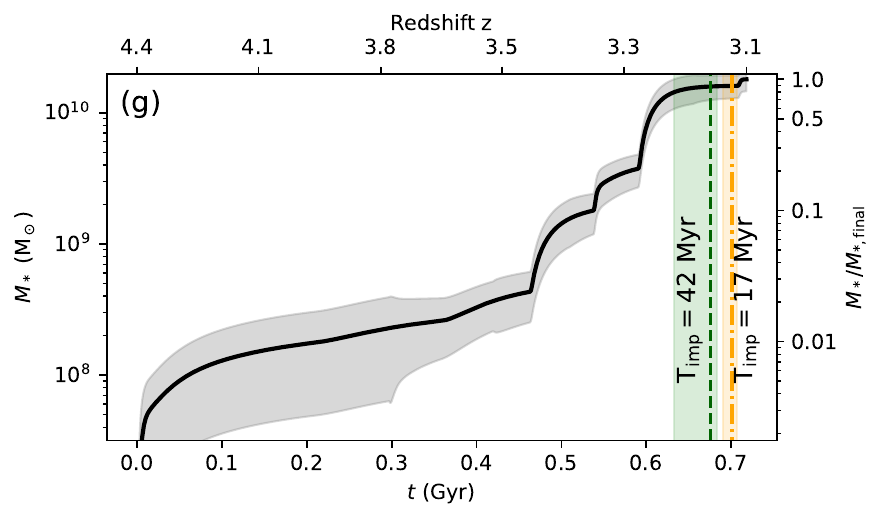}}

   \caption{\textsc{synthesizer-AGN} analysis results. Panels ({\it a})–({\it c}): stellar mass, fraction of the total stellar mass formed within the last 50~Myr, and fraction formed earlier, derived from \textsc{synthesizer-AGN}. Panels ({\it d})–({\it f}): star formation rate (SFR) and specific star formation rate (sSFR) over the last 50~Myr, and sSFR over the last 100~Myr. The stellar mass map (panel {\it a}) is shown in logarithmic scale, while all other maps are in linear scale. Panels ({\it b}) and ({\it c}) illustrate the relative contributions of recent and older stellar populations. All maps include contours of the [O III] line flux, tracing the ring and nuclear emission regions; the white diamond in panel ({\it a}) marks the mass barycentre. Panel ({\it g}) shows the global mass assembly history of the system \GS derived from \textsc{synthesizer-AGN}, where the grey shaded region display 1$\sigma$ uncertainties and the vertical lines mark the estimated epoch of the collision with the intruder (Sect.~\ref{sec:discussion}). 
   }
   \label{fig:stellarpropertiesSYN}%
   \end{figure}

The stellar mass map derived from \textsc{synthesizer-AGN} (Fig.~\ref{fig:stellarpropertiesSYN}, panel {\it a}) reveals an elongated, bar-like structure extending along the north–south direction and encompassing the northern and southern clumps of the ring. Hence, this structure is notably asymmetric with respect to the nucleus, deviating from the morphology of well-developed, dynamically settled bars observed in the local Universe.


We estimate a total stellar mass of $M_\star \simeq 1.7 \times 10^{10}$~\Msun for \GS (consistent with the SED-based value in \citealt{Pacifici2016}) and $M_\star \simeq 1 \times 10^{8}$~\Msun for the intruder. These values are obtained integrating over spaxel-by-spaxel measurements reported in the figure. 
Interestingly, the stellar mass barycentre (white diamond in the map) is offset by $\sim 0.8$~kpc northward from the gas kinematic centre (Sect.~\ref{sec:kinandfluxdistribution}). 
The innermost nuclear region (within $r = 0.2\arcsec$) contains approximately 15\% of the total stellar mass, while about half of the stellar mass resides along the extended bar-like structure (including the nuclear regions). 

The relatively shallow stellar mass concentration and the spatial offset of the mass barycentre suggest that the passage of the intruder has perturbed the inner potential of \GS, consistent with expectations for a recent collisional encounter (e.g. \citealt{Fiacconi2012}).




%

   \begin{figure*}[h]
   \centering
   \begin{minipage}{0.7\textwidth}
   \includegraphics[width=0.99\textwidth, trim=0mm 1mm 1mm 1mm, clip]{{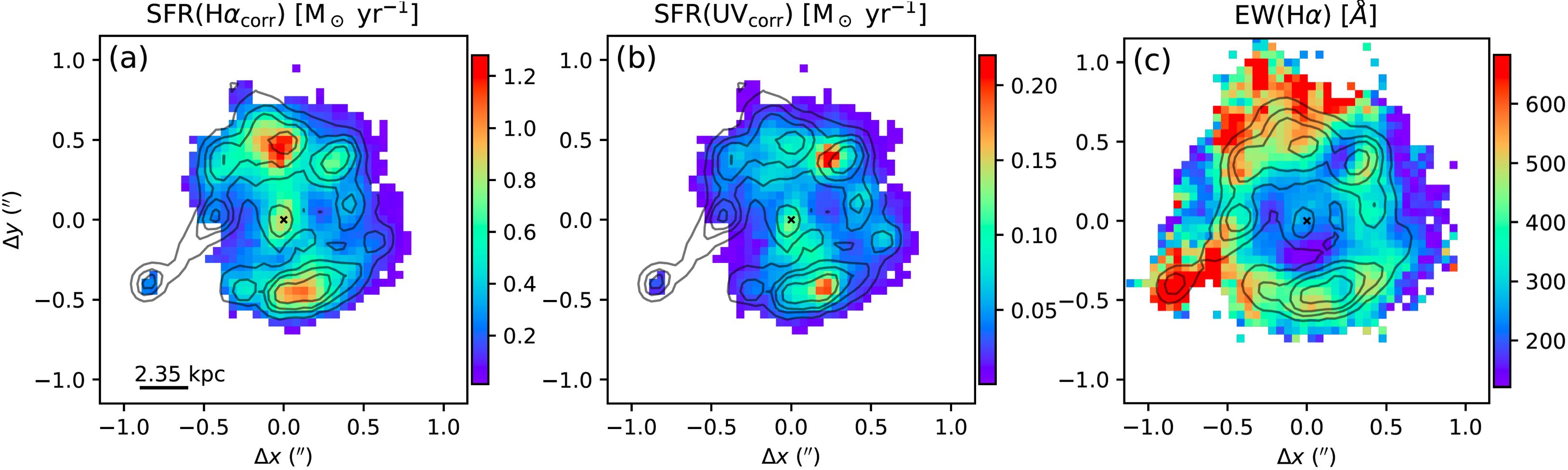}}
   \end{minipage}
   \hfill
 . \begin{minipage}{0.28\textwidth}
   \caption{Maps of the SFR derived from H$\alpha$ emission (panel {\it a}) and rest-frame UV continuum at $\sim 2300$~\AA\ (panel~{\it b}), both corrected for extinction (see Sect.~\ref{sec:sfr}), and of the H$\alpha$ equivalent width (panel {\it c}). Contours show the \oiii flux distribution. }
    \label{fig:SFRdiagnostics}%
    \end{minipage}
    \end{figure*}

\subsection{SFR from SED, UV and \ha emission}\label{sec:sfr}

The UV continuum emission is thought to trace star formation on time scales up to 100 Myr, while \ha is only sensitive to $\lesssim 10$~Myr time scales (\citealt{Kennicutt2012}).
From the single-spaxel fluxes, corrected for extinction using the $<E(B-V)>$ map shown in Fig.~\ref{fig:dustextinction} (bottom-right), we derive spatially-integrated ${\rm SFR(H\alpha)} \simeq 230$~\Msunyr\ and ${\rm SFR(UV)} \simeq 30$~\Msunyr for \GS; 
the intruder accounts for $\lesssim 1$\% of the total measurements. The significantly higher \ha-based value indicates a very recent and intense burst of star formation. 
SED fitting from R100 data provides a consistent value: ${\rm SFR(SED)} \simeq 40$~\Msunyr\ for \GS, and $\approx0.3$~\Msunyr\ for the intruder over the last 50~Myr. This time scale for the SED results is used to provide a better comparison with \ha and UV SFR tracers, sensitive to 10 and 100~Myr scales respectively (two times higher SED-based values are obtained over the last 25~Myr). 

Spatially, the SFR maps from H$\alpha$ and UV (Fig.~\ref{fig:SFRdiagnostics}, panels {\it a} and {\it b}) as well as from SED fitting (Fig.~\ref{fig:stellarpropertiesSYN}, panel {\it d}), all show that the strongest activity occurs mostly along the ring. 
The spatial distribution of the \ha equivalent width (EW), shown in Fig.~\ref{fig:SFRdiagnostics} (panel {\it c}), allows us to take a step further. Since EW(\ha) traces the ratio between current star formation (via the \ha line) and past star formation (via the adjacent continuum), higher values correspond to younger stellar populations (e.g. \citealt{Levesque2013, Reddy2018,Parlanti2025ring}).
In \GS, EW(\ha) peaks along the ring and immediate outer regions (especially in the north), and declines toward the centre; this confirms that the youngest stars are located in the ring region. Within the ring, regions of highest SFR(UV) correspond to relatively lower EW(\ha), suggesting a range of stellar ages among the star-forming clumps. Moreover, an increase of EW(\ha) toward the outer edge of the ring hints at a propagation of star formation to larger radii, or alternatively a radial decrease of the stellar continuum. 

\subsection{Mass assembly history}\label{sec:sfh}

The spatially integrated stellar mass assembly history derived from \textsc{synthesizer-AGN} and shown in Fig.~\ref{fig:stellarpropertiesSYN} (panel {\it g}) reveals a long phase of relatively modest star formation activity, followed by a sharp and rapid rise of mass assembly during the past few hundred Myr. This star formation enhancement may result from the early interaction phase and a later head-on collision between \GS and the intruder, consistent with predictions from simulations (e.g. \citealt{RodriguezGomez2016}). A comparable, though more temporally concentrated, burst is also recovered in the integrated stellar mass assembly history derived with \textsc{pPXF} (Fig.~\ref{fig:SFHppxf}, left), confirming the robustness of this result across different modelling approaches. 

The spatially resolved maps further support this interpretation. The specific SFR (sSFR = SFR/M$_*$) in  Fig.~\ref{fig:stellarpropertiesSYN} (panel {\it e}) closely mirrors the EW(H$\alpha$) distribution, both peaking along the ring.
Likewise, the map of the stellar mass fraction formed in the last 50~Myr (Fig.~\ref{fig:stellarpropertiesSYN}, panel {\it b}) demonstrates that the bulk of the most recent star formation is confined to the ring, with little activity in the central regions.
Together, these results provide evidence that the ongoing burst was likely initiated by the collision, consistent with a propagating star-formation wave triggered by the passage of the intruder through the disc.
The corresponding increase in recent stellar mass fraction, up to $\sim$10–20\% over the last few tens of Myr, also emerges from the \textsc{pPXF} and \texttt{CIGALE} analysis (see Appendix~\ref{app:ppxf} and \ref{app:cigale}). 




\section{Discussion}\label{sec:discussion}

Galactic rings are prominent structures whose formation could be attributed to several distinct mechanisms, each influencing the morphology, kinematics, and stellar and gas properties of the host galaxy and the ring itself. The primary formation processes include, in addition to collisional interactions (e.g. \citealt{Mihos1994}), resonant interactions (e.g. \citealt{Herrera-Endoqui2015}) and accretion from high-angular momentum cosmic-web streams (e.g. \citealt{Genzel2008,Finkelman2011, Dekel2020}). In this section, we show that \GS properties are most likely compatible with a collisional-induced ring scenario, on the basis of an examination of timescales, the resulting  physical, chemical, and kinematical structure characteristics of the ionised gas, the properties of the stellar populations, and the environmental conditions under which the ring formed.

\subsection{Evidence for a collisional origin}

Several independent lines of evidence indicate that the ring in \GS was produced by a recent collision with the nearby companion (the intruder). They include: the young stellar populations in the ring, its asymmetric and clumpy morphology, strong radial motions, tidal features, and the close kinematic association with a nearby companion. These aspects are discussed in detail below. 

In the framework of a collisional origin, the simple geometry of the ring, combined with the kinematics inferred in Sect.~\ref{sec:kinandfluxdistribution}, allows us to reconstruct the main features of the interaction. 
Two alternative methods can be used to estimate the interaction timeline. Assuming a uniform expansion velocity of the density wave, the timescale after the collision can be estimated as $T_{\rm imp} = r_{\rm ring} / v_{\rm exp}$, where $r_{\rm ring}$ is the ring radius and $v_{\rm exp}$ is the expansion velocity. For an expansion velocity equal to the radial velocity $v_{\rm rad} \approx 200$~\kms inferred from \barolo (Sect. \ref{sec:barolo}), we infer a timescale after the impact of $T_{\rm imp} = 17_{-7}^{+12}$~Myr (orange vertical line in Fig.~\ref{fig:stellarpropertiesSYN}, panel {\it g}, and Fig.~\ref{fig:SFHppxf}, left panel). This confidence interval is indicative, derived assuming typical 30\% uncertainties on the velocity and 20\% on the spatial extent of the ring. Moreover, we note that this $T_{\rm imp}$ estimate should be considered as a first-order approximation, since it assumes that the radial velocity component derived from \barolo 
accurately traces the density wave expansion velocity (e.g. \citealt{Toomre1978, Vorobyov2003}).

Alternatively, the timescale since the impact can be estimated based on the distance of the intruder from the main galaxy disk. If we assume a perpendicular passage of the intruder through the disc, we can deproject the observed distance using the inclination angle of the disc. Taking a projected separation of $d_p = 6.6$~kpc, and assuming an inclination  $i = 20^\circ$ (Sect.~\ref{sec:barolo}), the deprojected distance is $d_{\rm 3D} = d_p / sin(i) = 19_{-3}^{+11}$~kpc. Adopting a relative velocity of $425~$\kms (deprojected for the same inclination), the time since the collision is then estimated as $T_{\rm imp} = 42_{-8}^{+43}$~Myr (green vertical line in Fig.~\ref{fig:stellarpropertiesSYN}, panel {\it g}). The quoted ranges reflect the propagation of $1\sigma$ confidence intervals in inclination and projected distance (10$^\circ-25^\circ$ and 6.4 -- 6.8~kpc, respectively).
Taking into account the uncertainties in the derived quantities and the assumptions involved in the geometry and relative velocity, and that the two methods provide comparable results, we adopt an approximate timescale since the impact of $\approx 50$~Myr.

Comparable timescales are measured in nearby collisional ring galaxies. For example, \cite{Conn2016} derive an expansion age of $\sim 150$~Myr for the collisional ring in AM1354-250, based on an observed expansion velocity of 70~\kms (see also \citealt{Wallin1995}). Studies of the Cartwheel galaxy derive a wide range of ring ages from 70 to 720~Myr depending on tracer and ring location (\citealt{Amram1998, Higdon2015}), for an expanding velocity of $\sim 70$~\kms. At higher redshift, \citet{Yuan2020NatAs} inferred $> 40$~Myr for R5519 at $z \sim 2$, based on projected distances and $v_{\rm exp} = 230$~\kms.

The stellar mass assembly history derived from our \textsc{synthesizer-AGN} analysis provides further support for a recent, collision-driven origin of the ring. We find that roughly 10--20\% of the total stellar mass within the ring formed during the last few tens of Myr, comparable to the estimated time since the impact. This implies a powerful, short-lived burst of star formation, consistent with the compression of the gaseous disc by an expanding density wave following the passage of the intruder. Similarly, a high fraction of newly formed stars has been reported in the outer ring of the Cartwheel galaxy (\citealt{Ditrani2024}), where stellar populations are almost entirely dominated by young stars. The substantial recent mass growth in the ring of \GS\ is difficult to reconcile with a resonance scenario, where star formation proceeds more gradually and continuously over longer timescales (see e.g. \citealt{Costantin2023} and references therein).
Moreover, the clumpy, mildly asymmetric morphology of the ring in \GS is consistent with collisional rings and inconsistent with the smoother, symmetric structures expected in resonant rings (\citealt{Elmegreen2006}). 

While the stellar population analysis independently supports a recent, collision-triggered burst, the exact timing of the rise in stellar mass assembly derived from \textsc{synthesizer-AGN} does not coincide perfectly with the impact timescale estimated from the kinematics, whereas the \textsc{pPXF} and \texttt{CIGALE} results yield a more compatible temporal alignment (Fig.~\ref{fig:SFHppxf} and Table~\ref{tab:cigale}). Such differences reflect intrinsic uncertainties in reconstructing the detailed evolutionary history of complex (high-$z$) systems, and the comparison between the three independent methods suggests that systematic uncertainties on the temporal axis are unlikely to be smaller than a few tens of Myr. Moreover, the gravitational interaction between \GS and the companion starts before this collision, and this could have triggered star formation at $t < T_{imp}$.  Nonetheless, all methods consistently point to a recent rapid episode of mass assembly.

Finally, the presence of significant radial expansion velocities ($\sim 200$~\kms) further supports a collisional scenario. 
Moreover, the identification of the intruder galaxy at a small separation, with clear kinematic offsets ($\sim 425$~\kms), and additional tidal features and bridge-like structures connecting the galaxies strongly argue for a recent interaction between \GS and the intruder. 

Altogether, the observed properties of \GS (young stellar ages in the ring, comparable timescales of the impact and the stellar age in the ring, mildly asymmetric and clumpy morphology, strong radial motions, and the presence of a close interacting companion) converge towards a collisional origin as the most plausible explanation.

\subsection{Alternative scenario: Accretion from cosmic-web streams}  
In a cosmological framework, galaxies grow primarily through the accretion of gas from the surrounding cosmic web. 
\citet{Dekel2020} demonstrated that such accretion, associated with wet mergers in $\sim40\%$ of their simulated cases,  can trigger a compaction phase. This phase is then followed by inside-out quenching and a morphological transformation from a diffuse configuration into a compact central component surrounded by a large, gas-rich ring continuously fed by incoming cold streams (see their Fig. 4; see also \citealt{Zolotov2015}). These evolutionary processes typically occur on timescales of several hundred million years.
\citet{Dekel2020} identified possible observational detections in $z\sim 2$ compact galaxies with external rings presented, for instance, in \citet[see also \citealt{Tacchella2015} and \citealt{NestorShachar2025}]{Genzel2014quenching, Genzel2020}.


\GS shows signs of incipient inside-out quenching. This leaves open the possibility that the system is undergoing an early phase of the cold-accretion pathway proposed by \citet[][i.e. the `early post-compaction' phase]{Dekel2020}.
However, the other observational evidence, most notably the presence of a nearby companion, the disturbed kinematics, the presence of a bar-like structure, and the short timescale since the interaction matching the age of the stellar burst, are difficult to reconcile with the long-term evolution expected in the cosmic-web accretion scenario. 
We therefore consider this an interesting scenario for future investigation, but the available data favour a recent interaction as the primary driver of the ring in \GS.

\subsection{Alternative scenario: Resonant ring from a bar}  

Resonant rings form as a result of internal gravitational interactions within a galaxy hosting a bar structure. These rings are typically located at specific resonances where the orbital frequencies of stars and gas align with the pattern speed of the bar, leading to the accumulation of material. The formation timescale for resonant rings is closely tied to the development of the bar, which can occur over a few to up to several hundred million years (e.g. \citealt{SahaNaab2013,Bland-Hawthorn2024}). Once established, the rings can persist as long as the bar remains a dominant feature.
Therefore, resonant rings typically host a mixture of young and old stellar populations, and maintain smooth, rotation-supported morphologies.

In barred galaxies, the metallicity is expected to be higher along the bar structure and lower in the perpendicular direction (e.g. \citealt{Neumann2024}). 
A metallicity enhancement in the portion of the ring closest to the bar ends should also be expected in resonant rings (\citealt{Gajda2021}). 

The stellar mass distribution of \GS, derived from \textsc{synthesizer-AGN}, shows an elongated morphology extending roughly along the north–south direction. This could suggest that a bar-like structure is either present or in the process of forming. Recent observational studies have shown that bar-like instabilities may already occur in galaxies at $z > 2$ (e.g. \citealt{Costantin2023,Geron2025,Guo2025bars}), implying that the dynamical maturity required to develop such features can be achieved early in the Universe. Therefore, even if the ring in \GS was linked to a nascent bar, our results would still provide valuable insight into the early stages of bar-driven evolution and the coupling between internal and externally triggered instabilities in young discs.

However, several observational characteristics of \GS tend to argue against a purely resonant origin. 
The metallicty map (Fig.~\ref{fig:ZandU}, left) does not follow the expected pattern, lacking an enhanced value at the regions of the ring close to the bar-like structure, and also without clear differences along and across the structure. The lack of  ordered bar-driven motions along the structure (see residuals in Fig.~\ref{fig:3dbarolo}) also argue against this scenario. 
Moreover, the stellar mass residing along the extended structure significantly exceeds the standard bar-to-total stellar mass of low-$z$ galaxies ($\lesssim 20\%$; e.g. \citealt{Gadotti2011}). It is worth noting, however, that both observations and simulations of barred galaxies at $z \gtrsim 3$ remain extremely limited (see e.g. \citealt{Fragkoudi2025}), preventing a precise quantitative comparison of \GS with other systems.

Although we cannot completely exclude the possibility that the interaction triggered the formation of a bar (e.g. \citealt{Zhou2025bar}) and that the latter produced the ring only recently, such a sequence would require a finely tuned coincidence. Taking into account all the considerations discussed above, we therefore regard the collisional scenario as the most plausible explanation.

\subsection{Comparison with simulations of collisional ring galaxies}

Simulations of collisional ring galaxies show that axisymmetric encounters produce ring galaxies noticeably different with respect to those formed by off-centre interactions. Axisymmetric interactions result in a circular primary ring, often followed by smaller secondary rings. In contrast, those born from off-centre collisions show asymmetric rings and displaced nuclei (e.g. \citealt{Fiacconi2012, Chen2018ApJ864, Guo2022ApJ926}).
The almost-axisymmetric configuration in \GS may suggest an axisymmetric interaction, with relatively small impact parameter.

Another common prediction from simulations is the formation of a bar-like structure in interactions with non-zero impact parameter (e.g. \citealt{Fiacconi2012, Chen2018ApJ864}). \citet{Renaud2018} further showed that `spokes', filamentary features similar to those seen in the Cartwheel galaxy, may form as a result of gas inflow from the ring to the centre. In their model, star formation is initially concentrated in the ring and later propagates inward through the spokes, eventually reaching the nucleus. These spokes can efficiently channel gas and stars from the ring to the centre, dissolving the ring within $\sim 200$~Myr after the interaction. The bar-like structure observed in \GS could therefore be interpreted either as a structure formed by an off-centre impact, as proposed in \citet{Fiacconi2012}, or as the signature of spatially-unresolved spokes funnelling material and star formation from the ring to the inner regions, as shown in \citet{Renaud2018}.

Simulations also predict strongly disturbed gas distributions resulting from the interaction. The gaseous ring often appears broad, clumpy, and fragmented. The azimuthal velocity field is likewise perturbed, with deviations from rotational symmetry that are more pronounced in systems with lower bulge-to-disc ratios, particularly at large radii. A peak in the azimuthal velocity profile is typically observed at the ring radius. Vertical (non-planar) motions are also commonly generated, especially in inclined encounters. Such interactions may also produce thin gaseous tails along the intruder’s path and induce warps in the outer disc (e.g. \citealt{Fiacconi2012}).
These kinematic features (non-circular motions, vertical velocities, tails, and possible warping) may be consistent with the residuals observed in the \barolo modelling (Fig.~\ref{fig:3dbarolo}), and may serve as further evidence supporting a collisional origin for the ring in \GS.

A potential concern arises when comparing the stellar mass ratio between \GS and the intruder ($\sim$ 1:100) with those commonly adopted in simulations (e.g. 1:10 or higher in \citealt{Chen2018ApJ864,Renaud2018}). At face value, such a low mass ratio might appear inconsistent with the formation of a strong collisional ring, as very minor companions would behave more like long-lived satellites with long orbital decay timescales and a reduced ability to produce a ring. However, both simulations and observations show that post-collision mass ratios can differ substantially from the pre-impact values. In particular, \citet{Renaud2018} found that the intruder in their Cartwheel simulations loses a large fraction of its dynamical mass during the interaction, with much of it dispersed into tidal debris, consistent with the observational properties of the Cartwheel system itself. In addition, recent star formation in the target galaxy can bias the inferred stellar mass ratio toward smaller values (\citealt{Hernquist1993}). These effects imply that the current mass ratio of \GS and its intruder may not reflect the pre-collision mass ratio relevant for producing the ring. The difficulties in reconciling present-day stellar masses with the progenitor masses inferred from simulations therefore mirror those encountered in the prototypical Cartwheel galaxy, and do not argue against a collisional origin for the ring in \GS.


\section{Conclusions}\label{sec:conclusions}

In this work, we present \GS, the most distant ring galaxy to date, identified at a spectroscopic redshift of $z = 3.076$ using JWST/NIRSpec IFS observations. This discovery provides a unique high-redshift laboratory to investigate the formation mechanisms and evolutionary significance of ring galaxies in the early Universe.
Our detailed analysis of \GS's morphology, kinematics, stellar populations, and ionized gas properties strongly supports a collisional origin. 

\begin{itemize} 
\item Morphology: \GS exhibits a prominent, clumpy ring structure ($r_{\rm ring} \sim 3.5$~kpc) with a bright nuclear region and a bar-like feature. Faint tidal tails are also observed. This morphology, particularly the well-defined and clumpy ring, and the presence of a close companion connected to \GS with a bridge, is a hallmark of collisional ring galaxies formed by head-on interactions, as shown by simulations. The observed bar-like feature in GS18660 aligns with predictions from simulations for interactions with a non-zero (but still small) impact parameter. 
\item Kinematics: 
The companion is moving away from \GS with a significant velocity offset of $\sim 425$~\kms. It has a tail-like structure that points towards the nuclear regions of \GS. Kinematic analysis reveals a rotating disc component with an additional radial velocity of approximately $\sim 200$~\kms, consistent with an outward propagating density wave triggered by a collision.  
Simulations show that collisional ring galaxies have disturbed gas distributions and perturbed azimuthal velocity fields, with a peak at the ring radius, in addition to the radial planar motions, and also exhibit vertical velocities perpendicular to the disk. These disturbed kinematic features and potential disc warping may explain the residuals in our kinematic modelling with \barolo, and hence be consistent with a collisional origin for \GS's ring. 
\item Interaction Timescale: Our estimates of the timescale since impact, ranging from $\sim 17$ to $42$~Myr (with a combined confidence interval going from 10 to 85~Myr), are short and comparable to those derived for other nearby and $z=1-2$ collisional ring galaxies. 
\item  Stellar Populations: The ring hosts young stars formed as a result of the interaction (i.e. in the last few tens of Myr), while innermost regions show less prominent stellar activity. A radial increasing trend in star-formation activity is supported by a radial increasing trend in sSFR and \ha EWs. Roughly 10--20\% of the total stellar mass within the ring formed during the last few tens of Myr, comparable to the estimated time since the impact. 
\item ISM Properties: The ionised gas properties indicate ongoing star formation along the ring. Electron densities in \GS's ring are elevated (100--1000 cm$^{-3}$) compared to lower-$z$ analogues, possibly reflecting more compact and intense star-forming regions. Electron temperatures are also elevated but within expected ranges for high-$z$ star-forming regions. Gas metallicity is enhanced in the nuclear region but lower in the ring and connecting bridge structures, suggesting that gas in the ring was not previously enriched by older stellar generations, consistent with expectations for collisional ring galaxies and in line with our stellar population analysis results. Emission line ratios are consistent with stellar ionisation; the nuclear regions show lower \oiii/\hb and higher \nii/\ha compared to the ring, consistent with other nearby collisional ring galaxies. 
\item Environmental context: We identified a close companion, the intruder galaxy, moving away from \GS with a significant velocity offset of $\sim 425$~\kms. The presence of this companion and other two line emitters in the vicinity suggests that \GS resides within a small galaxy group, a common characteristic of collisional galaxies at lower redshifts and predicted by cosmological simulations. 
\end{itemize}
In conclusion, \GS may represent a rare and significant case of a high-redshift collisional ring galaxy. Its observed morphological and kinematic features, combined with its short interaction timescale, highly active star-forming ring, and distinct ISM properties, align remarkably well with numerical simulations of collisional ring formation. 

While the collected evidence strongly supports a collisional origin, we cannot fully exclude the possibility that \GS hosts a recently formed bar and that the observed ring corresponds to an early resonance feature. If confirmed, such a case would provide one of the most distant examples of bar-driven secular evolution, offering a rare window into the onset of bar formation and its dynamical impact on galaxy structure at $z > 3$. A conclusive settlement of these scenarios will require deeper follow-up observations aimed at characterising the stellar emission in the galaxy outskirts and constraining the faint, older stellar components. Deep JWST/NIRCam imaging in the rest-frame optical would be particularly valuable to map the underlying stellar continuum and reveal whether the bar-like structure represents a product of the recent collision.

Regardless of its precise ring formation mechanism, the discovery of \GS, empowered by JWST’s capabilities, advances our understanding of galaxy interactions and their role in shaping galaxy evolution in the early Universe.

\begin{acknowledgements}


We thank E. Di Teodoro, I. Shivaei, J. \'Alvarez-M\'arquez, A. Chakraborty, and G. Tozzi for valuable discussions.

MP, SA, LC, BRP, PPG, and CPJ acknowledge support from the research projects PID2021-127718NB-I00, PID2024-159902NA-I00, PID2024-158856NA-I00, and RYC2023-044853-I of the Spanish Ministry of Science and Innovation/State Agency of Research (MCIN/AEI/10.13039/501100011033) and FSE+.
The project that gave rise to these results received the support of a fellowship from the “la Caixa” Foundation (ID 100010434). The fellowship code is LCF/BQ/PR24/12050015.
IL acknowledges support from grant PRIN-MUR 2020ACSP5K\_002 financed by European Union – Next Generation EU.
RM acknowledges support by the Science and Technology Facilities Council (STFC), by the ERC Advanced Grant 695671 ``QUENCH'', and by the UKRI Frontier Research grant RISEandFALL; RM is further supported by a research professorship from the Royal Society.
AJB acknowledges funding from the ``First Galaxies'' Advanced Grant from the European Research Council (ERC) under the European Union’s Horizon 2020 research and innovation programme (Grant agreement No. 789056).
%
%
SC and GV acknowledge support by European Union’s HE ERC Starting Grant No. 101040227 - WINGS.
MP, GC and EB acknowledge the support of the INAF Large Grant 2022 "The metal circle: a new sharp view of the baryon cycle up to Cosmic Dawn with the latest generation IFU facilities". 
GC and EB also acknowledge the INAF GO grant ``A JWST/MIRI MIRACLE: Mid-IR Activity of Circumnuclear Line Emission''. EB acknowledges funding through the INAF ``Ricerca Fondamentale 2024'' programme (mini-grant 1.05.24.07.01).

H\"U acknowledges funding by the European Union (ERC APEX, 101164796). Views and opinions expressed are however those of the authors only and do not necessarily reflect those of the European Union or the European Research Council Executive Agency. Neither the European Union nor the granting authority can be held responsible for them.
This research made use of astrodendro, a Python package to compute dendrograms of Astronomical data (\url{http://www.dendrograms.org/})
\end{acknowledgements}

%
%

\bibliographystyle{aa}
\bibliography{ringpaper.bib}

\begin{appendix}

\section{HST images}\label{sec:aHST}
   \begin{figure}[!h]
   \centering
   \includegraphics[width=0.4\textwidth, trim=0mm 0mm 4mm 8mm, clip]{{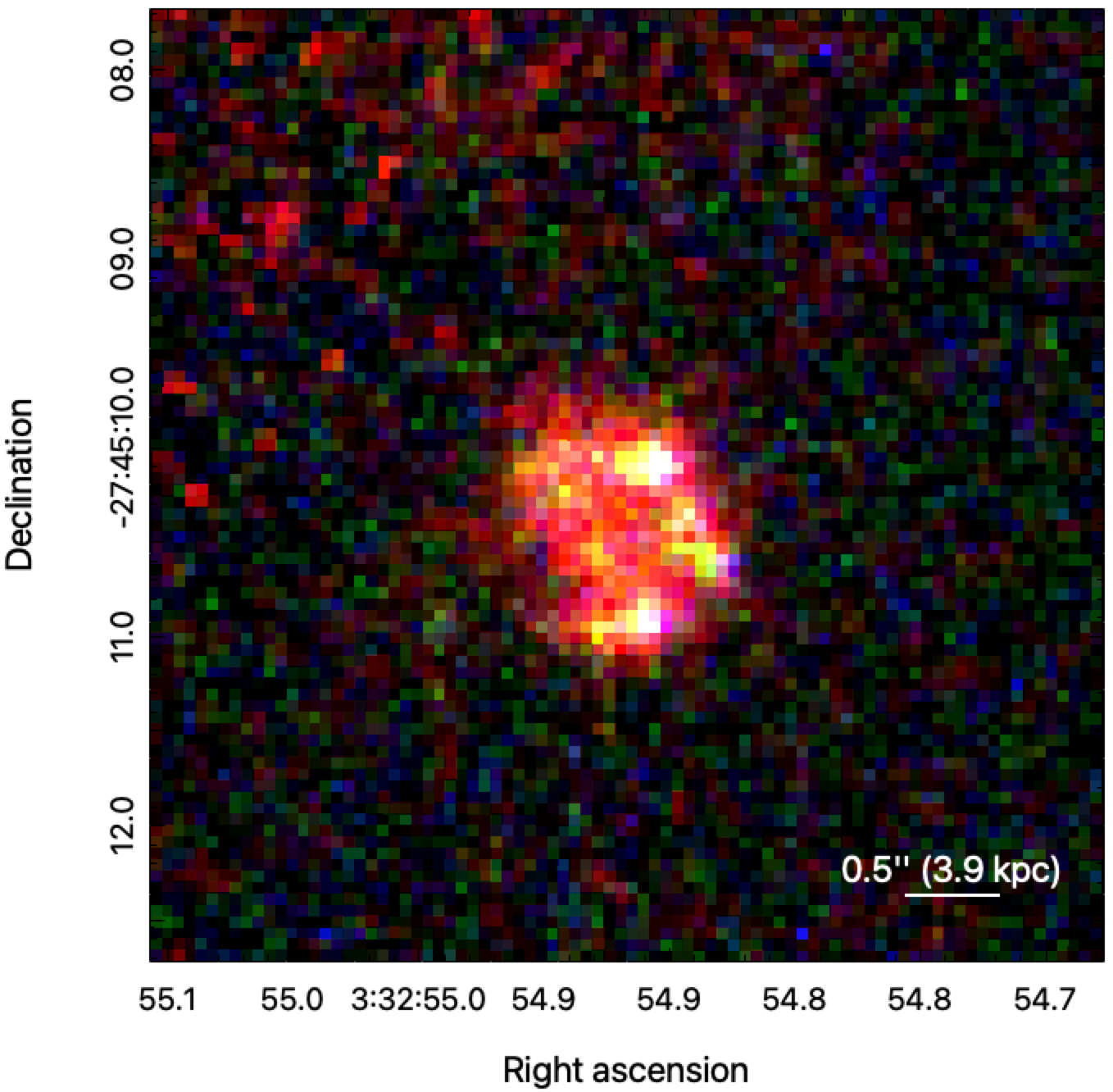}}
   \caption{RGB image of \GS from HST ACS-i, ACS-z, and WFC3-F160W (from Rainbow Cosmological Surveys database, \url{https://arcoirix.cab.inta-csic.es/Rainbow_navigator_public}).}
              \label{FigGam}%
    \end{figure}

\section{R100-R2700 velocity offset}\label{sec:R100offset}

   \begin{figure}[!h]
   \centering
    \includegraphics[width=0.45\textwidth, trim=0mm 2mm 0mm 2mm, clip]{{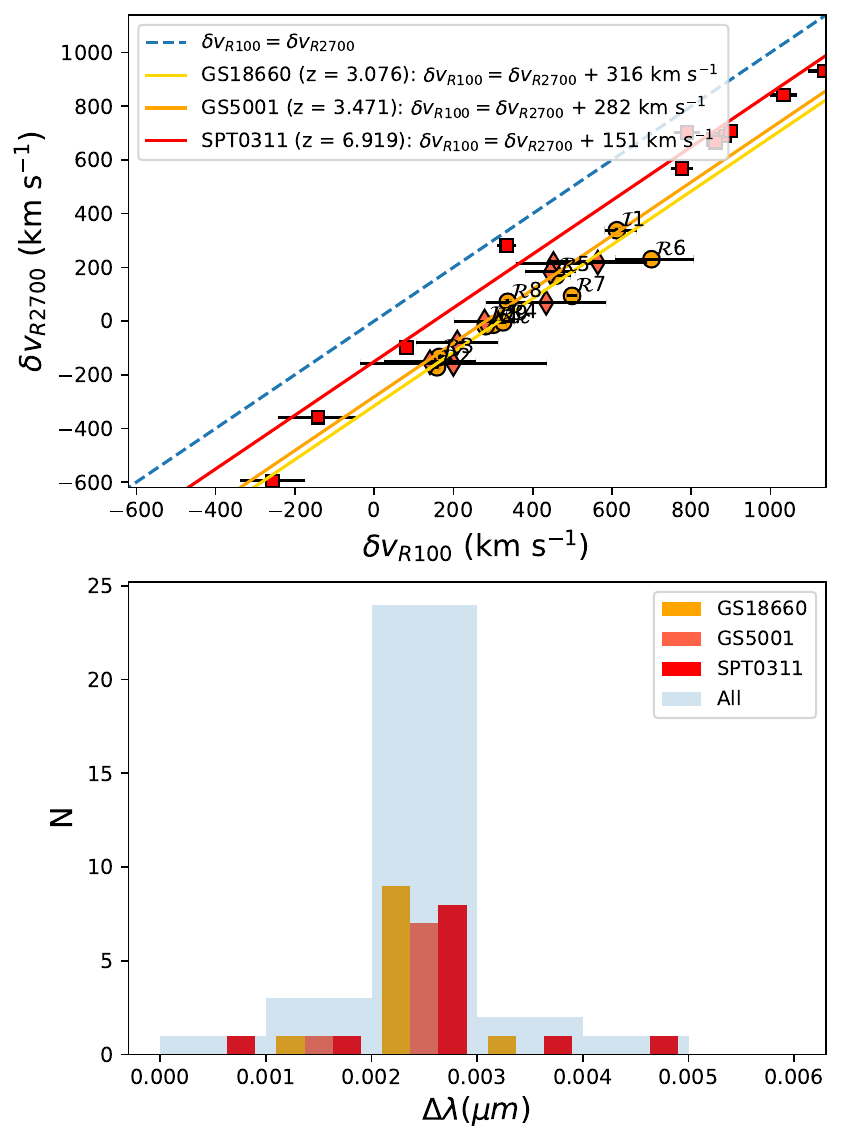}}

    \caption{Top panel: Comparison of observed velocities of emission lines in R100 spectra with those obtained from R2700 for the same spatial regions in \GS (yellow circles), GS5001 (orange diamonds, from \citealt{Lamperti2024}), and SPT0311-58 (red squares, \citealt{Arribas2024}). For \GS we assume $z = 3.07662$, for GS5001 $z = 3.47111$, for SPT0311-58 $z = 6.902$. The 1:1 relation is in blue, the best-fit relations for individual datasets are shown with different colours, as labelled. 
    Bottom panel: Distribution of $\Delta \lambda$ values (in observer-frame) inferred from the three datasets. }
    \label{fig:DVoffset}%
    \end{figure}

To investigate the wavelength calibration accuracy of the NIRSpec R100 data, we compared the observed velocities of emission lines in the low-resolution R100 spectra with those obtained from the high-resolution R2700 data for the same spatial regions.

We extracted integrated spectra from several bright star-forming regions within the \GS system, as defined with the dendrogram technique (Sect.~\ref{sec:results:ism}). We then fitted simultaneously all bright emission lines detected in the R100 and R2700 spectra (see Figs.~\ref{fig:spectraleaves} and \ref{fig:spectraleavesR100}). Across all regions, we consistently found that the emission lines in the R100 spectra are redshifted by $\sim 315$~\kms relative to R2700 data. This was quantified by fitting a linear relation of the form: $\delta v_{R100} = \delta v_{R2700} + B$ where $B$ is the offset. This result is consistent with the velocity offset measured from spectra integrated over the entire ring galaxy (Fig.\ref{fig:3coloursNIRSpec}). The comparison between the measurements from the different regions and the best-fit relation are reported in Fig.~\ref{fig:DVoffset}, left (yellow circles and solid line).

A comparable velocity offset is observed in the GA-NIFS target GS5001 \citep{Lamperti2024}, which contains multiple clumps and galaxies within the NIRSpec FOV at $z\sim3.47$. The velocity measurements for GS5001, shown as orange diamonds in Fig.~\ref{fig:DVoffset}, yield a slightly smaller but consistent offset of $\sim280$~\kms (orange best-fit line).
We also examined the GA-NIFS target SPT0311  (\citealt{Arribas2024}), also containing multiple galaxies at $z \sim 6.90$ within the NIRSpec FOV. These measurements, shown in red, display consistent but slightly smaller offsets ($\sim 151$~\kms, red best-fit line).

In all three systems, the velocity offset is uniform across the field, suggesting a systematic origin in the wavelength calibration of the R100 configuration. Notably, the observed offset decreases with redshift, from $\sim315$~\kms at $z \sim 3$ to $\sim150$~\kms at $z \sim 7$, consistent with a fixed wavelength offset in the observer frame.

To quantify this effect, we computed the correction required to adjust the wavelength axis of R100 cubes, assuming the offset can be absorbed in a modification of the \texttt{CRVAL3} FITS header keyword. For a given source, the corrected rest-frame wavelength can be computed as:
\begin{equation}
\lambda_{\rm new} = \lambda \times \left(1 - \frac{\Delta v}{c}\right),
\end{equation}
where $\Delta v = \delta v_{\mathrm{R100}} - \delta v_{\mathrm{R2700}}$ is the measured velocity offset. Equivalently, the correction to be applied to \texttt{CRVAL3} in the observer frame is:
\begin{equation}
\Delta \lambda_{\rm obs} = \texttt{CRVAL3} \times \left(\frac{\Delta v}{c}\right) \times (1 + z).
\end{equation}

Assuming a fixed wavelength offset $\Delta \lambda_{\rm obs} = \Delta \lambda_{\rm obs}^\prime$, this relation implies:
$\Delta v \times (1 + z) = \Delta v' \times (1 + z'),
$
where $\Delta \lambda_{\rm obs}$, $\Delta v$ and $z$ refer to one NIRSpec IFU dataset, and $\Delta \lambda_{\rm obs}^\prime$, $\Delta v'$, $z'$ to another. 

We used this scaling to calculate the product $\Delta v \times (1 + z)$ for all clumps and galaxies in the three datasets. The resulting distribution of this quantity, which corresponds to the observer-frame wavelength shift $\Delta\lambda_{\rm obs}$, is shown in Fig.~\ref{fig:DVoffset} (bottom panel). The figure displays a narrow distribution, with median $\Delta \lambda = 24_{-3}^{+5}$~\AA.

These results indicate that a systematic wavelength calibration offset affects the R100 mode in NIRSpec IFS observations. We recommend applying the correction described above when using R100 data for velocity measurements in the absence of an absolute wavelength reference.




\section{Additional data reduction procedures}\label{app:extraDR}

{\bf Astrometric registration.}
Astrometric registration is usually applied to match NIRSpec data with external imaging or multi-wavelength observations (e.g. \citealt{Perna2024arp}). However, since in this work we do not combine NIRSpec with other datasets, no astrometric correction is applied.

\noindent{\bf Spurious feature at 1.09 $\mu$m.}
Some NIRSpec datasets are affected by an artificial emission-like feature at $\sim 1.09~\mu$m (observer frame), extending across the full FOV. While this feature is not present in the \GS observations, it has been identified in earlier NIRSpec R100 (e.g. \citealt{Arribas2024}), R1000 (e.g. Costantin et al., in prep.), and R2700 data (\citealt{Cresci2023}). A commonly adopted correction involves subtracting an in-field background spectrum. It is important to emphasise that, even in cases where this spurious feature is not visible in the raw science data, the sflat used in the reduction pipeline may still exhibit this artifact. This can introduce residual effects in the shape of the continuum around 1.09~$\mu$m, and should be considered when interpreting the spectral energy distribution in this wavelength region (e.g. $\approx 2670~\AA$ rest-frame in \GS spectra).

\noindent{\bf Wiggles correction.}
Previous JWST/NIRSpec IFS observations revealed the presence of sinusoidal artefacts (‘wiggles’) in single-spaxel spectra of R2700 cubes of bright sources (e.g. \citealt{Marshall2023, Ulivi2025, Bertola2025}). These wiggles, which are caused by under-sampling of the PSF, can be modelled and subtracted with the routine presented in \citet[][see also \citealt{Dumont2025}]{Perna2023}. 


\section{Line emitter}
\begin{figure}[!h]
   \centering
    \includegraphics[width=0.4\textwidth]{{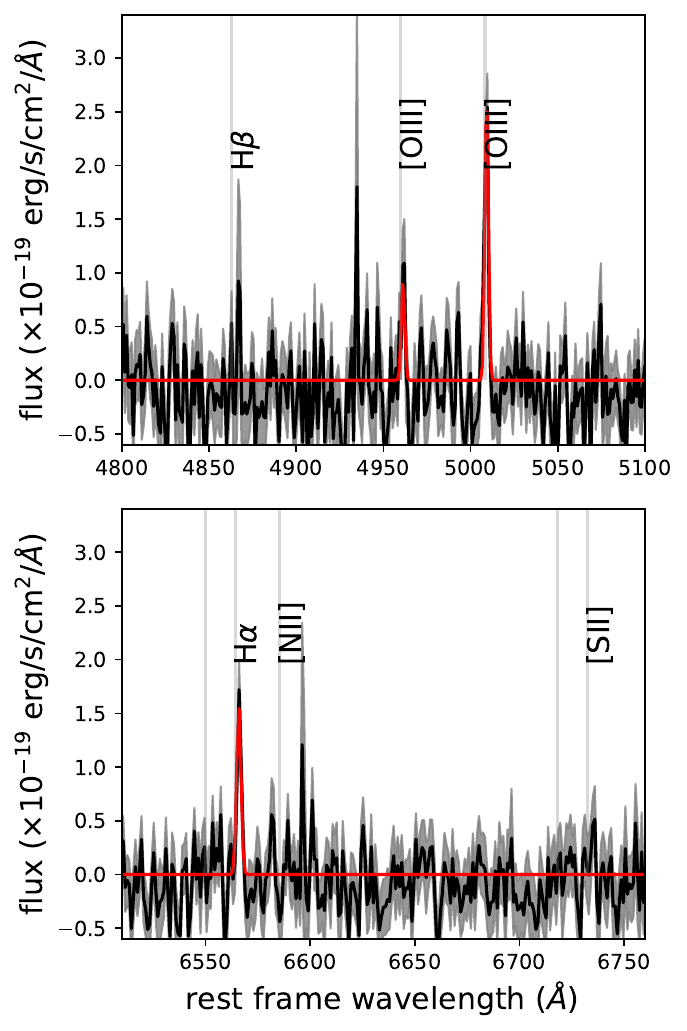}}

    \caption{Integrated R2700 spectrum of the line emitter (black curve), with best-fit results (red). The spectrum is extracted within $r=0.15\arcsec$; 3$\sigma$ uncertainties are shown in grey. The main emission lines are marked with grey vertical lines and labelled.}
    \label{fig:integratedspectraLINEEMITTER}%
    \end{figure}

The R2700 integrated spectrum of the line emitter labelled in Fig.~\ref{fig:3coloursNIRSpec} is reported in Fig.~\ref{fig:integratedspectraLINEEMITTER}.

\section{R100 and R2700 integrated spectra}

\begin{figure*}[!p]
\centering
\includegraphics[width=1.01\textwidth]{{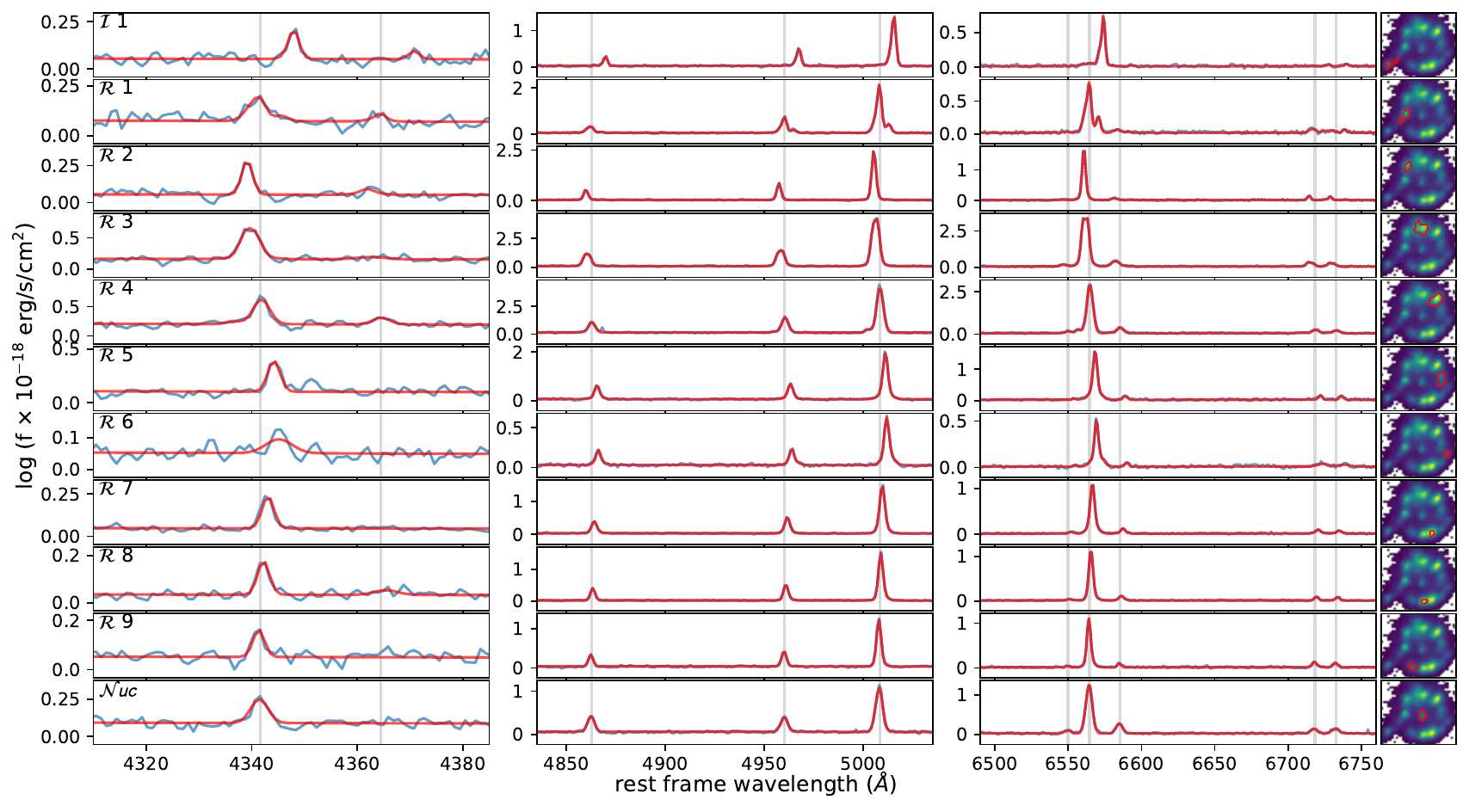}}
\caption{R2700 spectra (blue) with best-fit results (red) extracted from the different structures of \GS and the intruder. These structures were identified as the smallest unresolved line-emitting elements with the dendrogram technique. Each region is highlighted in the right panel with a red contour over the \ha emission map. Vertical lines in the spectra mark the position of \hg and \oiii 4363 (left), \hb and \oiii 4960, 5008 (centre), \ha, \nii and \sii doublets (right).See Fig. \ref{fig:spectraleavesR100} for R100 spectra extracted from the same regions.}
          \label{fig:spectraleaves}%
\end{figure*}

\begin{figure*}
\centering
\includegraphics[width=1.01\textwidth]{{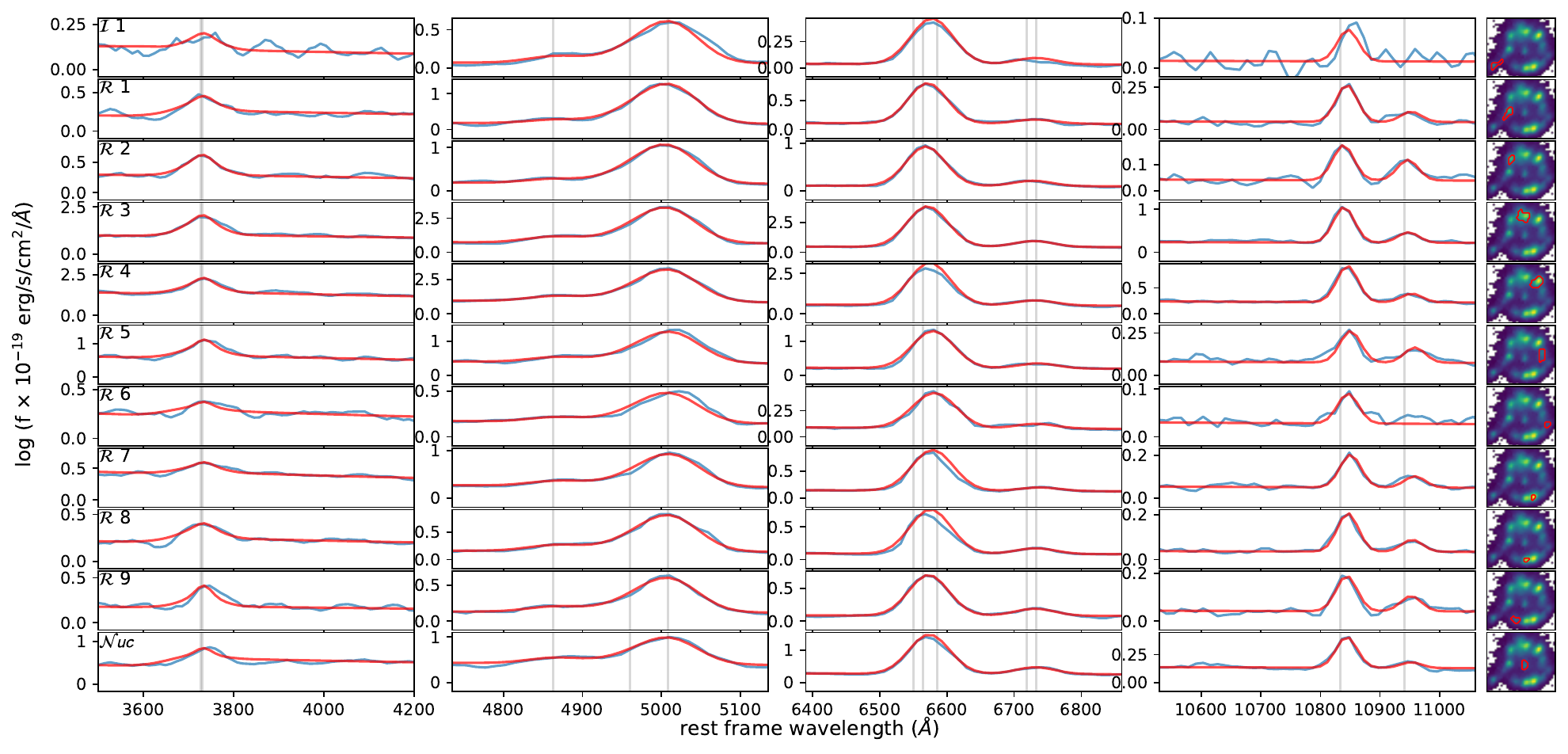}}
\caption{R100 spectra (blue) with best-fit results (red) extracted from the different structures of \GS and the intruder. These structures were identified as the smallest unresolved line-emitting elements with the dendrogram technique. Each region is highlighted in the right panel with a red contour over the \ha emission map. See Fig. \ref{fig:spectraleaves} for R2700 spectra extracted from the same regions.}
          \label{fig:spectraleavesR100}%
\end{figure*}

R100 and R2700 integrated spectra from dendgrogam leaves introduced in Sect.~\ref{sec:results:ism} are reported in Figs.~\ref{fig:spectraleavesR100} and \ref{fig:spectraleaves}.
The best-fit kinematic and physical properties derived for each region are summarised in Table~\ref{tab:integratedproperties}.

%

\begin{table*}
\tabcolsep 1.5pt 
\centering
\caption{Spatially integrated properties for individual clumps.}%
\begin{tabular}{|l cc cc ccccc c c c|}
\hline

Region  & $\delta v$ & $\sigma_v$ &  log(L\oiii) & log(L\ha) & A$_{\rm V}$ & $n_e$ & T$_e$ & Z$_{R3,~N2}$ & Z$_{R2,~R3,~R23,~N2}$ & log($U$) & SFR & EW(\ha)\\
  & (\kms) & (\kms) & (\ergs) & (\ergs) & (mag) & (cm$^{-3}$) & (10$^4$~K) & \multicolumn{2} {c} {(12+log(O/H))} &  & (\Msunyr) & ($\AA$)\\
\hline\hline

$\mathcal{I}1$ & $415_{-6}^{+12}$ & $50_{21}^{+3}$ & $41.54_{-0.01}^{+0.01}$ & $41.31_{-0.01}^{+0.01}$ & $0.53_{-0.06}^{+0.05}$ & -- & $1.9_{-0.1}^{+0.1}$ & -- &$<8.00$ & $-3.10_{-0.50}^{+0.30}$ & $1.22_{-0.30}^{+0.30}$ & $780_{-30}^{+100}$\\

$\mathcal{I}1b$ & $150_{-160}^{+10}$ & $310_{-170}^{+20}$ & $40.60_{-0.20}^{+0.20}$ & $40.80_{-0.20}^{+0.20}$ &  -- & --& -- & -- & --& -- & -- & --\\

\hline

$\mathcal{R}1$ & $-50_{-1}^{+1}$ & $94_{-1}^{+1}$ & $41.78_{-0.01}^{+0.01}$ & $41.48_{-0.01}^{+0.01}$ & $0.32_{-0.03}^{+0.04}$ & $135_{-40}^{+20}$ & $1.4\pm 0.1$& $8.30\pm0.08$ &$8.30\pm0.08$ &$-2.58_{-0.11}^{+0.06}$ & $2.0_{-0.1}^{+0.1}$ & $470_{-22}^{+15}$\\

$\mathcal{R}1b$ & $275_{-1}^{+1}$ & $50_{-1}^{+2}$ & $41.00_{-0.02}^{+0.02}$ & $40.88_{-0.02}^{+0.02}$ & $>0.97$ & -- & -- & -- & -- & -- & $>3.4$ & --\\

\hline

$\mathcal{R}2$ & $-166_{-6}^{+15}$ & $105_{-20}^{+20}$ & $41.78_{-0.03}^{0.03}$ & $41.69_{-0.09}^{+0.09}$ & $0.74_{-0.15}^{+0.06}$ & $1100_{-300}^{+400}$ & $1.9_{-0.3}^{+0.1}$ & $8.42\pm0.06$ &$8.43\pm0.06$ & $-2.88_{-0.05}^{+0.13}$ & $5.4_{-1.1}^{+0.3}$ & $570_{-30}^{+20}$\\

\hline

$\mathcal{R}3$ & $-127_{-1}^{+1}$ & $115_{-5}^{+1}$ &$42.25_{-0.01}^{0.01}$ &$42.28_{-0.01}^{+0.01}$ & $1.3_{0.1}^{0.1}$ & $360_{-60}^{+80}$ & $1.3_{-0.1}^{+0.5}$ & $8.37\pm0.05$ &$8.39\pm 0.05$& $-2.83_{-0.03}^{+0.04}$ & $29.0_{-3.0}^{+1.6}$ & $370_{-10}^{+10}$\\

\hline

$\mathcal{R}4$ & $9_{-1}^{+1}$ & $88_{-1}^{+1}$ & $42.18_{-0.01}^{+0.01}$ & $42.12_{-0.01}^{+0.01}$ & $0.9_{-0.1}^{+0.1}$ & $310_{-90}^{+60}$ & $1.7_{-0.1}^{+0.1}$ & $8.35\pm0.04$ &$8.36\pm0.04$ &$-2.70_{-0.02}^{+0.05}$ & $15.7_{-0.8}^{+0.7}$ & $210_{-25}^{+10}$\\

$\mathcal{R}4b$ & $-350_{-5}^{+2}$ & $64_{-1}^{+1}$ & $40.96_{-0.02}^{+0.02}$ & $40.91_{-0.02}^{+0.02}$ & -- & -- & -- & -- & --& -- & -- & --\\

\hline
$\mathcal{R}5$ & $170_{-1}^{+1}$ & $120_{-3}^{+4}$ & $41.79_{-0.03}^{+0.03}$ & $41.80_{-0.02}^{+0.02}$ & $0.7_{-0.1}^{+0.1}$ & $450_{-140}^{+220}$ & $>1.9$ & $8.40\pm0.07$ &$8.41\pm0.08$& $-2.84_{-0.08}^{+0.10}$ & $4.2_{-0.7}^{+0.4}$ & $214_{-30}^{+10}$\\

\hline
$\mathcal{R}6$ & $228_{-2}^{+2}$ & $128_{-6}^{+3}$ & $41.30_{-0.05}^{+0.05}$ &$41.30_{-0.05}^{+0.05}$ & $0.5_{-0.1}^{+0.1}$ & $310_{-150}^{+250}$ & $>2.4$ & $8.40\pm0.10$ &$8.39\pm0.10$& $-2.87_{-0.10}^{+0.07}$ & $1.2_{-0.2}^{+0.1}$ & $230_{-30}^{+50}$\\

\hline

$\mathcal{R}7$ & $81_{-2}^{+2}$ & $92_{-2}^{+3}$ & $41.65_{-0.01}^{+0.01}$ & $41.63_{-0.03}^{+0.03}$ &  $0.9_{-0.1}^{+0.1}$ & $220_{-75}^{+80}$ & $>1.5$ &$8.36\pm0.04$ &$8.36\pm0.05$& $-2.53_{-0.05}^{+0.09}$ & $4.2_{0.3}^{0.3}$ & $320_{-30}^{+20}$\\

\hline

$\mathcal{R}8$ & $43_{-3}^{+1}$ & $84_{-6}^{+5}$ & $41.61_{-0.03}^{+0.03}$ &$41.59_{-0.02}^{+0.02}$ & $0.9_{-0.1}^{+0.1}$ & $1140_{-190}^{+280}$ & $>1.9$ &$8.39\pm 0.06$ &$8.40\pm 0.07$& $-2.72_{-0.03}^{+0.06}$ & $7.8_{-1.3}^{+2.8}$ & $230_{-20}^{+20}$\\

\hline

$\mathcal{R}9$ & $-17_{-3}^{+1}$ & $64_{-4}^{+7}$ & $41.50_{-0.04}^{+0.04}$ & $41.54_{-0.02}^{+0.02}$ & $1.5_{-0.3}^{+0.1}$ & $490_{-300}^{+180}$ &$>2.3$ & $8.39\pm 0.07$& $8.40\pm 0.06$& $-2.91_{-0.04}^{+0.03}$ & $3.7_{-0.5}^{+0.4}$ & $460_{-20}^{+40}$\\

\hline

$\mathcal{N}uc$ & $-19_{-1}^{+1}$ &  $93_{-1}^{+1}$ & $41.60_{-0.05}^{+0.05}$ & $41.77_{-0.02}^{+0.02}$& $1.3_{-0.1}^{+0.1}$ & $930_{-280}^{+320}$ &  $>1.7$ &$8.45\pm0.04$ &$8.45\pm0.04$& $-2.89_{-0.09}^{+0.07}$ & $9.5_{-0.9}^{+1.1}$ & $210_{-20}^{+20}$\\

\hline

\end{tabular} 
\tablefoot{\GS properties from R2700 and R100 spectra integrated over the dendrogram leaves shown in Fig.~\ref{fig:spectraleaves}. When multiple kinematic components are detected, the secondary component is labelled with the suffix {\it b}. The metallicity measurements are inferred combining R2 and R23 (from R100, correcting for extinction inferred from high-resolution spectra), and R3 and N2 (from R2700); the ionisation potential log($U$) is derived from R100 spectra; all other measurements in the table are obtained from R2700, less affected by emission line blending. Luminosities are not corrected for extinction. Extinction measured from \ha/\hb (Sect.~\ref{sec:extinction}). Electron density and temperature inferred from \sii and \oiii diagnostics, respectively (Sects. \ref{sec:electrondensity}, \ref{sec:electrontemperature}). Metallicities inferred from the line ratios mentioned in the table (Sect.~\ref{sec:metallicity}). Ionisation potential from \sii/\siii (Sect. \ref{sec:ionisation}). SFR from extinction-corrected \ha fluxes (Sect.~\ref{sec:sfr}). }   
\label{tab:integratedproperties}
\end{table*}

\section{pPXF stellar population synthesis fit }\label{app:ppxf}

We analyse each single-spaxel spectrum with the Penalized Pixel-Fitting algorithm (\textsc{pPXF} ; \citealt{Cappellari2017,Cappellari2023}) and derive the non-parametric SFH modelling both the stellar and gas components. We fit the spectra with the BPASS library 
\citep[version 2.2.1; ][]{Eldridge2017, Stanway2018} 
and the \cite {Chabrier2003} IMF. 
We consider models in a 2D age-metallicity logarithmic grid, spanning 34 age bins from 1 Myr to the age of the Universe at $z = 3.08$ (plus a buffer of 200~Myr) and 10 [M/H] bins from $-1.75$ to 0~dex. 

To retrieve the SFH of \GS, we followed the methodology
described in \citet{Ikhsanova2025}, but (1) using \texttt{degree}=5 and \texttt{mdegree}=9 for fixing the kinematics; (2) limiting the regularization 
to a maximum value of \texttt{regul} = 100.
Uncertainties are estimated with a bootstrap
analysis in the distribution of the
mass weights of each Voronoi bin \citep[see e.g. ][]{Kacharov2018}.

   \begin{figure*}
   \centering
   \includegraphics[width=0.99\textwidth, trim=0mm 0mm 0mm 0mm,clip]{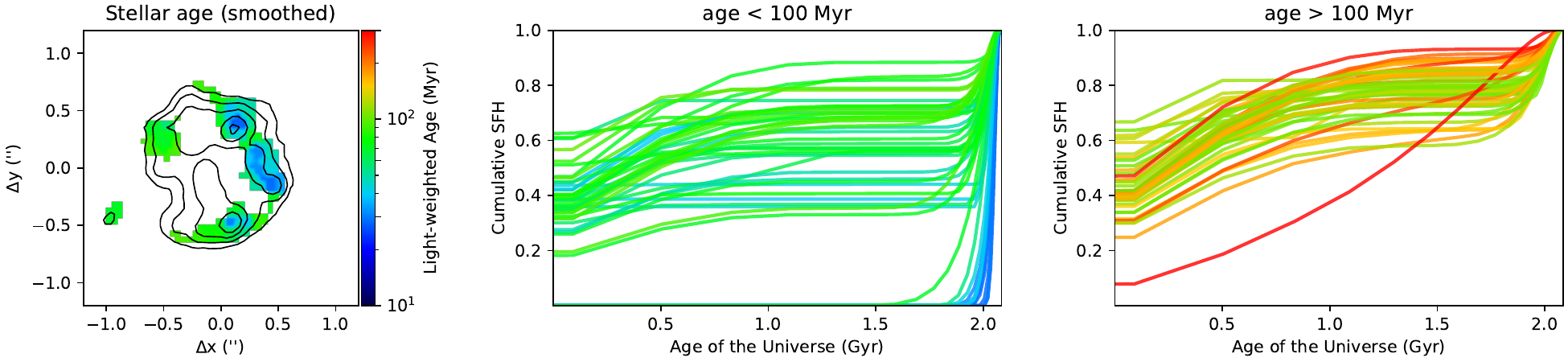}
   \caption{Stellar age distribution and SFHs from \textsc{pPXF} analysis. Each SFH line is obtained from individual Voronoi bin; the SFHs are divided in two panels according to the light-weighted age of the Voronoi bins, for visual purposes. Regions with youngest light-weighted ages are localized in the ring, and show the most recent and most rapid episodes of star formation. Regions with oldest ages are not represented in the map. }
              \label{fig:SFHppxf}%
    \end{figure*}

   \begin{figure*}
   \centering
   \includegraphics[width=0.99\textwidth, trim=0mm 0mm 0mm 0mm,clip]{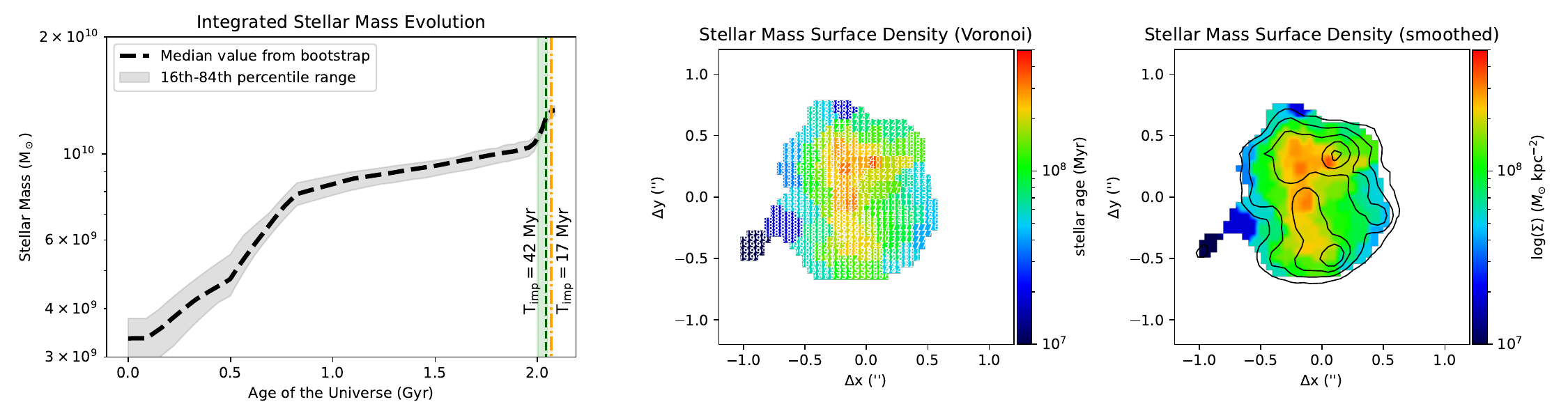}
   \caption{Global mass assembly history (left panel), and stellar mass surface density (middle and right panels) derived from \textsc{pPXF} analysis. }
              \label{fig:SFHppxf}%
    \end{figure*}

\section{SFH with CIGALE }\label{app:cigale}

We perform an SED fitting using the spectro-photometric version of \texttt{CIGALE} (\citealt{Noll2009,Boquien2019,burgarella25}). The redshift used in the analysis is the one derived spectroscopically in this work. We make different assumptions to model the SFH of the system, with the aim of constraining the age of the most recent stellar population. The best results were selected by minimising the $\chi$$^{2}$ value and ensuring best agreement with the observed spectra. 

The most satisfactory results were obtained when considering two components: an old population of 200 Myr and a constant young burst ranging from 0 to 100~Myr. We used the \cite {Bruzual2003} stellar population models and the \cite {Chabrier2003} IMF. Nebular continuum and emission lines were modelled using an electron density of 100~cm$^{-3}$, an ionization parameter (logU) ranging from -3.2 to -2.6, and gas metallicity between 0.3 and 1 Z$_{\odot}$. The colour excess of the nebular lines ranges from 0.01 to 1 and we assumed E(B-V) $_{nebular}$\,=\,E(B-V)$_{stellar}$. Dust emission was modelled using the \cite{Draine2007} models. 

\begin{figure*}
   \centering
    \includegraphics[width=0.45\textwidth, trim=0mm 2mm 0mm 2mm, clip]
    {{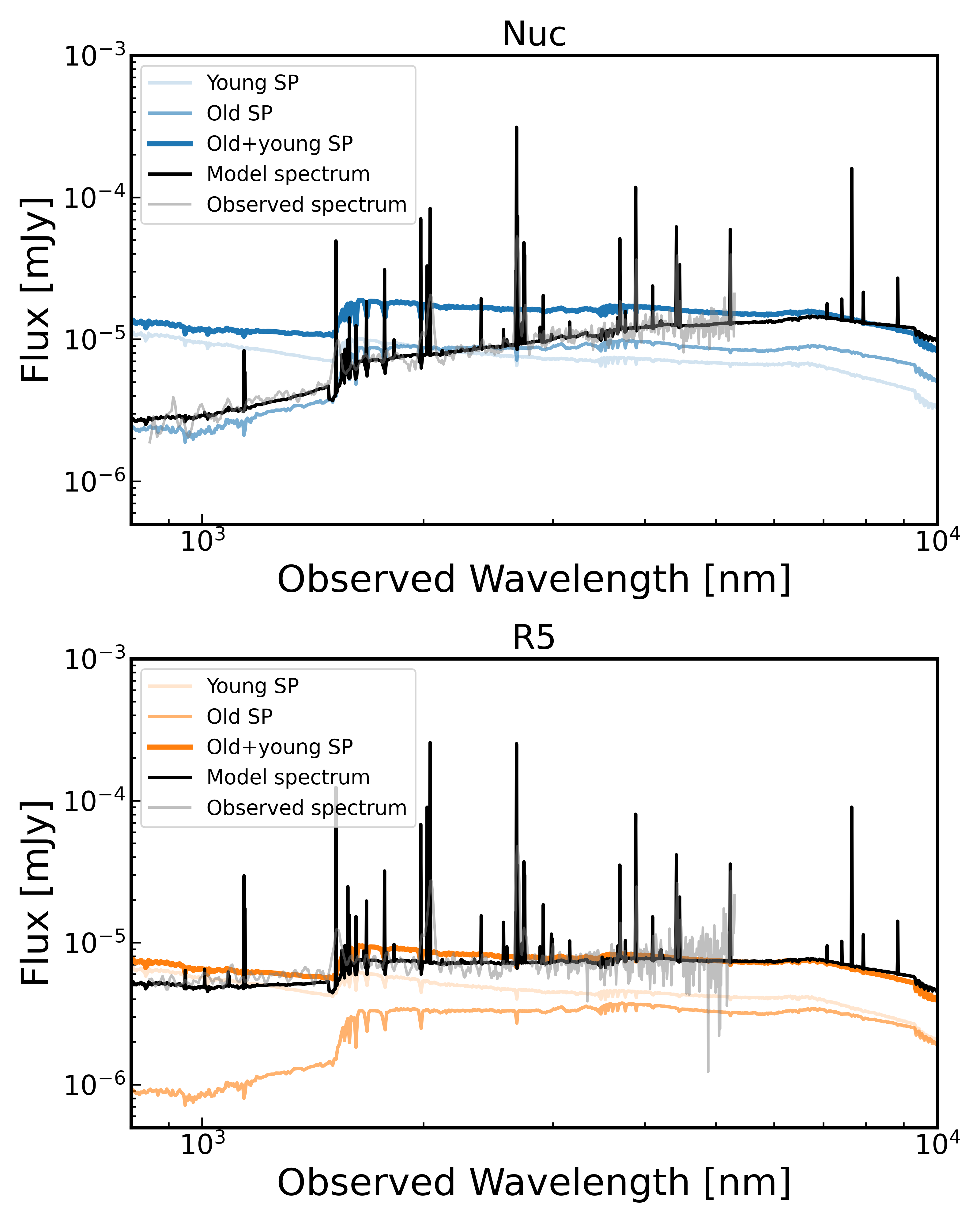}}
   \caption{ Best-fit model of the nuclear and $\mathcal{R}5$ integrated regions (according to Fig.~\ref{fig:spectraleavesR100}). It is also represented the contribution to the SED of the models of the young, old and old+young unattenuated stellar spectra.}
              \label{fig:cigale_nuc_r5}%
    \end{figure*}

The ages of the individual clumps are reported in Table \ref{tab:cigale}, and the best-fit spectra for the regions of the nucleus and $\mathcal{R}5$ (according to Fig.~\ref{fig:spectraleaves}) are shown in Fig.~\ref{fig:cigale_nuc_r5}.
The derived ages of the clumps in the ring yield an average value of $75_{-46}^{14}$~Myr, consistent with the steep rise in the stellar mass assembly history inferred from both \textsc{sythesizer-AGN} and \textsc{pPXF}. Although uncertainties remain significant, the overall consistency between the \texttt{CIGALE}-derived burst ages and the impact timescale estimated in Sect.~\ref{sec:discussion} supports a coherent picture in which the recent starburst was triggered by the collision. The fraction of stellar mass formed in the recent episode, as inferred from \texttt{CIGALE}, is also in good agreement with that obtained from the other two codes.

\begin{table}[h]

\centering
\caption{Results of the SED fitting obtained with \texttt{CIGALE}. Burst ages, mass fractions, and H$\alpha$ equivalent widths.}
\begin{tabular}{|lccc|}
\hline
Region & Age burst & f$_\text{burst}$ & EW(\ha) \\
       & (Myr)     &                 & ($\AA$) \\

\hline
\hline
$\mathcal{I}1$  & 17 $\pm$ 21 & 0.16 $\pm$ 0.27 & 553 $\pm$ 53 \\
$\mathcal{R}1$  & 9 $\pm$ 16  & 0.04 $\pm$ 0.08 & 398 $\pm$ 21 \\
$\mathcal{R}2$  & 81 $\pm$ 16 & 0.58 $\pm$ 0.16 & 455 $\pm$ 21 \\
$\mathcal{R}3$  & 75 $\pm$ 18 & 0.43 $\pm$ 0.32 & 394 $\pm$ 46 \\
$\mathcal{R}4$  & 96 $\pm$ 8  & 0.30 $\pm$ 0.02 & 296 $\pm$ 13 \\
$\mathcal{R}5$   & 92 $\pm$ 12 & 0.28 $\pm$ 0.05 & 277 $\pm$ 17 \\
$\mathcal{R}6$  & 92 $\pm$ 9  & 0.19 $\pm$ 0.02 & 219 $\pm$ 10 \\
$\mathcal{R}7$   & 73 $\pm$ 10 & 0.18 $\pm$ 0.03 & 250 $\pm$ 8 \\
$\mathcal{R}8$   & 81 $\pm$ 19 & 0.52 $\pm$ 0.25 & 388 $\pm$ 24 \\
$\mathcal{R}9$   & 58 $\pm$ 21 & 0.30 $\pm$ 0.24 & 398 $\pm$ 25 \\
$\mathcal{N}uc$  & 80 $\pm$ 15 & 0.18 $\pm$ 0.03 & 262 $\pm$ 19 \\
\hline
\end{tabular}
\tablefoot{\GS properties obtained from the prism spectra integrated over the dendrogram leaves shown in Fig.~\ref{fig:spectraleavesR100}. The fraction of the burst represents the mass fraction of stars formed in the recent burst relative to the total stellar mass of the galaxy, as labelled.}
\label{tab:cigale}
\end{table}

\end{appendix}
\end{document}